\documentclass[numberedappendix]{emulateapj}
\usepackage{amsmath}
\usepackage[dvipsnames]{xcolor}
\usepackage{verbatim}
\usepackage{multirow}

\slugcomment{accepted by ApJ; December 11, 2017}

\newcommand{\G}{\Gamma}

\newcommand{\sT}{\sigma_{\rm T}}
\newcommand{\p}{^\prime}
\newcommand{\e}{\epsilon}
\newcommand{\g}{\gamma}
\newcommand{\gp}{\gamma^{\prime}}

\newcommand{\ep}{\epsilon^\prime}

\newcommand{\dD}{\delta_{\rm D}}

\newcommand{\psim}{\lower.5ex\hbox{$\; \buildrel \propto \over\sim \;$}}
\newcommand{\lbar}{\lower.0ex\hbox{$\; \buildrel
{\lower0.0ex \hbox{-}} \over\lambda  \;$}}

\newcommand{\cm}{\mathrm{cm}}
\newcommand{\km}{\mathrm{km}}
\newcommand{\erg}{\mathrm{erg}}

\newcommand{\GeV}{\mathrm{GeV}}

\newcommand{\s}{\mathrm{s}}

\newcommand{\pc}{\mathrm{pc}}

\newcommand{\Mpc}{\mathrm{Mpc}}

\def\Ngreen{N_e}
\newcommand{\angstrom}{\text{\normalfont\AA}}

\shorttitle{Electron Acceleration in Blazars}
\shortauthors{Lewis, Finke, \& Becker}

\begin{document}
\title{A Steady-State Spectral Model For Electron Acceleration and Cooling in Blazar Jets: Application to 3C 279}

\author{Tiffany R.\ Lewis}
\affil{	Department of Physics \& Astronomy, MSN 3F3, 
	George Mason University, 4400 University Drive, Fairfax, VA 22030}
\email{tlewis13@masonlive.gmu.edu}

\author{Justin D.\ Finke}
\affil{U.S.\ Naval Research Laboratory, Code 7653, 4555 Overlook Ave.\ SW,
        Washington, DC,
        20375-5352}
\email{justin.finke@nrl.navy.mil}

\author{Peter A.\ Becker}
\affil{	Department of Physics \& Astronomy, MSN 3F3,
	George Mason University, 4400 University Drive, Fairfax, VA 22030}
\email{pbecker@gmu.edu}

\begin{abstract}

We introduce a new theoretical model to describe the emitting region in a blazar jet.  We assume a one-zone leptonic picture, and construct the particle transport equation for a plasma blob experiencing low-energy, monoenergetic particle injection, energy dependent particle escape, shock acceleration, adiabatic expansion, stochastic acceleration, synchrotron radiation, and external Compton radiation from the dust torus and broad line region.  We demonstrate that a one-zone leptonic model is able to explain the IR though $\g$-ray spectrum for 3C 279 in 2008-2009. We determine that the broad-line region seed photons cannot be adequately described by a single average distribution, but rather we find that a stratified broad line region provides an improvement in the estimation of the distance of the emitting region from the black hole.   We calculate that the jet is not always in equipartition between the particles and magnetic field, and find that stochastic acceleration provides more energy to the particles than does shock acceleration, where the latter is also overshadowed by adiabatic losses.   We further introduce a novel technique to implement numerical boundary conditions and determine the global normalization for the electron distribution, based on analysis of stiff ordinary differential equations. Our astrophysical results are compared with those obtained by previous authors.

\end{abstract}

\keywords{quasars: general --- radiation mechanisms: nonthermal ---galaxies: active --- galaxies: jets --- (galaxies:) quasars: individual (3C 279) --- radiative transfer }

\section{Introduction}
\label{intro}


Blazars are a subclass of active galactic nuclei with bipolar relativistic jets aligned with our line of sight.  The jet orientation, combined with Doppler boosting, causes the jet emission to dominate observations at nearly all wavelengths.  The jet emission is highly variable, with variability timescales as short as \, $\sim\,$minutes in some cases \citep[e.g.][]{aharonian07,aleksic11,ackermann16}. Blazars are divided into two categories; flat spectrum radio quasars (FSRQs) have strong broad emission lines, and BL Lac objects have weak or absent lines.  

\subsection{Motivation}

The broadband $\nu F_\nu$ spectral energy distributions (SEDs) of blazars have two broad components (``bumps''): a low-frequency one peaking between IR and X-rays, and a high-frequency one peaking in the $\g$-rays.  The lower-frequency component is almost certainly synchrotron emission from nonthermal electrons (including positrons) accelerated in the jet.  The radio emission in blazars is thought to be generated outside of the primary ($\gamma$-ray) emission region, due to the superposition of synchrotron spectra from a range of jet locations where the optical depth $\tau_{\rm opt}\approx1$ \citep[e.g.][]{blandford79, konigl81}, and cannot be formed by the same component as the low-frequency peak emission process because the former must be self-absorbed.  The origin of the higher-energy component is a matter of debate.  It could be due to Compton scattering of synchrotron photons by the same nonthermal electrons that produce the synchrotron (synchrotron self-Compton, or SSC), which is thought to produce the $\g$-rays in high-peaked BL Lac objects \citep[e.g.,][]{maraschi92,dermer93,sikora94,bloom96,blazejowski00}.  Or it could be due to the Compton scattering of lower-energy seed photons external to the jet (external Compton or EC) from an accretion disk \citep[e.g.,][]{dermer92,dermer93}, broad-line region (BLR)\citep[e.g.,][]{sikora94,blandford95,ghisellini96}, or dust torus \citep[e.g.,][]{kataoka99,blazejowski00} by the same electrons producing the synchrotron.  The seed photon source is closely tied to the location in the jet of the $\g$-ray emitting region. For an emitting region at distances $r\la0.01\ \pc$ from the black hole, scattering of disk photons dominates; for an emitting region in the range $0.01 \la r \la 0.1\ \pc$, scattering of BLR photons dominates; and for distances $0.1 \la r \la 1\ \pc$, scattering of dust torus photons dominates. 

Alternatively, the $\g$-rays could be produced by hadronic processes:  proton synchrotron \citep[e.g.,][]{aharonian00,muecke01,mucke03,reimer04} or by the decay products of proton-photon interactions \citep[e.g.,][]{sikora87,mannheim92,mannheim93,protheroe95} with protons co-accelerated in the jet with the electrons, and the synchrotron emission of the resulting particles produced by these hadronic processes.  Hadronic processes for $\g$-ray production are in some cases disfavored due to excessive energetics \citep[e.g.,][]{boett13,zdziarski15,petropoulou16}, leaving Compton scattering as the most likely source for $\g$-rays. 


The electron acceleration mechanism operating during the observed blazar flares is currently not well understood  \citep[e.g.][]{madejski16,romero17}. The emitting electrons are probably energized by some combination of shock acceleration \citep[e.g.,][]{summerlin12,marscher14}, stochastic scattering \citep[e.g.,][]{katarzynski06,lefa11,asano15,baring17}, or electrostatic acceleration due to magnetic reconnection \citep[e.g.,][]{giannios09,giannios13,petropoulou16_reconn,sironi16}.  Often, rather than modeling the acceleration mechanism, a functional form for the emitting electron distribution is assumed, and theoretical SEDs are calculated from this \citep[e.g.,][]{tavecchio98,bednarek97,bednarek99,finke08_SSC,dermer09,hayash12}.

We endeavor to improve on these types of models by determining the electron distribution using a self-consistent framework, in which we treat particle acceleration, escape, and energy losses by solving a Fokker-Planck equation.  We include diffusive shock and stochastic acceleration, adiabatic and radiative losses, described by the full Compton cross-section.  In particular, including the full Compton cross-section is more physically realistic, and can be important in the case where the high-energy SED peak is brighter compared with the low-energy peak.  The Klein-Nishina (KN) turnover can cause inefficient losses at high energies, leading to a harder electron distribution, and hence harder resulting synchrotron and Compton spectra \citep[e.g.][]{dermer02_KN,moderski05}.  \citet{dermer14} studied the use of a log-parabola electron distribution for modeling blazars, in particular for modeling SEDs of the FSRQ 3C 279, in an equipartition framework.  These authors had difficulty explaining the $\g$-ray emission detected from the SEDs by the {\it Fermi} Large Area Telescope (LAT) at $\ga 5\ \GeV$.  All of the models we present here are able to explain $\ge 5$ GeV $\g$-rays detected by the {\it Fermi}-LAT.  Other authors \citep[e.g.,][]{diltz14,asano14,asano15} have performed time-dependent modeling of several blazar flares, including second order particle acceleration, and energy losses described by the full Compton cross-section.  We have created a similar model, although we use a time-independent, steady-state analysis for simplicity.  However, we include a more complete physical modeling of the BLR \citep{finke16} and scattering of dust torus photons, both being important seed photon sources for $\g$-ray production.  


\citet{dermer14} applied their modeling technique to the well-studied FSRQ 3C 279.  This was the first blazar detected by EGRET on board the {\em Compton Gamma-Ray Observatory} in 1991 \citep{hartman92}, and only the second blazar detected in $\g$-rays after 3C 273 by COS B \citep{swanenburg78}.  In 2008, the {\it Fermi}-LAT began collecting data for 3C 279 as part of its standard survey mode, providing regular monitoring of this source and reinvigorating multiwavelength efforts.  The first multiwavelength campaign for 3C 279  found a large optical polarization swing associated with a $\g$-ray outburst detected by {\it Fermi}-LAT \citep{abdo10_3c279}.   \citet{hayash12} performed further analysis on several spectral epochs.  They modeled the broadband SEDs with broken power-law electron distributions and two scenarios:  a propagating emission region model, and a model with two distinct synchrotron emission regions to account for the high- and low-energy features simultaneously.  The sequence of SEDs provided by \citet{hayash12} are relatively complete, in both wavelength and time, and provide an excellent resource for SED modeling.

The FSRQ 3C 279 has a redshift $z=0.536$, giving it a luminosity distance $d_L= 3100\ \Mpc$ in a cosmology where $H_0=70\ \km\ \s^{-1}\ \Mpc^{-1}$, $\Omega_m=0.3$, $\Omega_\Lambda=0.7$.  We use this distance in all of our relevant calculations.

\subsection{Physical Picture}


Our goal in this paper is to develop a new one-zone leptonic model, in which the jet emission is assumed to come predominantly from one emitting region--a homogenous spherical ``blob.''  We therefore neglect the radio emission, which is emitted elsewhere \citep[e.g.][]{blandford79, konigl81}. The propagation angle relative to the line of sight, $\theta$, introduces a Doppler factor $\dD = [\G(1-\beta\cos\theta)]^{-1}$.  We will here assume that $\dD=\G$.  The blob contains nonthermal electrons, randomly oriented (isotropic) in the frame co-moving with the blob (the ``co-moving frame'').  We assume the co-moving blob radius $R\p_b$ is constrained by the light crossing timescale $t_{v}$, so that $R\p_b\la c\dD t_{v}/(1+z)$.  The blob moves at a highly relativistic speed, $v=\beta c$, giving it a bulk Lorentz factor $\G = (1-\beta^2)^{-1/2}$. The electrons are accelerated by diffusive shock and stochastic processes inside the blob, which also contains a tangled, turbulent magnetic field.  The blob may also contain protons, but these are radiatively unimportant in our model. 

The blob is embedded in a jet launched from a supermassive black hole (BH) at the center of the active galaxy.  Matter accretes onto the black hole, creating a thermal accretion disk with temperature profile and emission given by \citet{shakura73}.  The disk's emission sometimes appears as a ``blue bump'' in the optical/UV spectra of quasars.  We will refer to the frame stationary with respect to the BH as the ``stationary frame''.

The optical spectra of AGN often contain broad emission lines, indicating the presence of highly ionized, fast-moving gas in a broad line region (BLR).  Reverberation mapping indicates the BLR gas is traveling in Keplerian orbits around the BH, and different lines emit primarily at different distances from the BH \citep[e.g.,][]{peterson99}.  The BLR absorbs disk photons and re-emits them as line radiation (the eponymous broad lines). Some of these BLR photons make their way to the jet blob, where they are Compton scattered by the nonthermal electrons there.  The exact geometry of the BLR is not well known.  In blazar spectral modeling, it is sometimes assumed that Lyman-$\alpha$ will be the dominant emission source from the BLR, and that other emission lines can be neglected; \citep[e.g.,][]{moderski03,dermer14} or that the whole BLR spectrum can be represented by a thermal distribution at a single radius \citep[e.g.,][]{ghisellini98,tavecchio08,ghisellini08}.

  \citet{joshi14} used 35 lines in their BLR model, assuming all of the line emission was emitted throughout a single shell regardless of line.   In contrast, \citet{finke16} created a stratified BLR model, with 26 different lines emitting at different radii, in a spherical shell or flattened configuration with the BH at the center \citep[see also][]{abolmasov17}, and we adopt the same picture here.  The radii of the different line sources were based on a SDSS composite quasar spectrum \citep{vanden01}.  \citet{finke16} provided formulae for  computing the accurate Compton-scattering spectra from these lines, taking into account the complicated geometry of the BLR.  We model the stratification of the BLR in a simplified fashion, using an isotropic approximation for the Compton scattering formulae, and allow the energy density from a particular monochromatic line to decrease outside of the radius where that line is emitted.  Thus, in this picture, the most prominent emission line for Compton scattering, out of the 26 considered here, depends on the location of the blob.

We also include a dust torus in our model, which absorbs and re-emits a fraction of the disk photons. The infrared emission from the torus is modeled similarly to a BLR line, as a radiation field isotropic in the stationary frame that is monochromatic, with photon energy given by the peak of a blackbody spectrum.  Some subset of the infrared dust photons become seed photons for Compton scattering by electrons in the blob.


Our model is based on a quantitative description of the basic physical processes experienced by the electrons contained in a magnetically-confined plasma blob in the emitting region of the blazar jet.  We are especially interested in formulating a self-consistent correspondence between the underlying electron distribution and the observed spectral distribution, where practicable.  Specifically, the electrons experience first- and second-order Fermi acceleration, Bohm diffusion (energy dependent escape), adiabatic losses, synchrotron losses, and external Compton (EC) losses due to interactions with seed radiation from the disk, dust torus, and the BLR. The spectral model also includes synchrotron self-absorption (SSA) and SSC processes.  SSC losses were neglected in the solutions to the electron transport equation since this would make the equation nonlinear \citep[e.g.,][]{zacharias12,zacharias12_EC,zacharias14}.  However, these losses are unlikely to have a strong effect, since EC losses will dominate.  We consider the full Compton cross-section in our transport equation solution \citep[e.g.,][]{boett97} and spectral calculations \citep[e.g.,][]{jones68,blumen70},  where the KN regime becomes important at higher energies.  


\subsection{Previous Work}

We compare our steady-state model to multiwavelength SED data for 3C 279, originally published in \citet{hayash12}. Some of the light curves for these SEDs were first published in \citet{abdo10_3c279}, but not all of the SEDs we use were incuded, and they did not model the SEDs in that paper.  \citet{hayash12} use a broken power law or double broken power law electron distribution as necessary, with both one- and two-zone models.  They used the X-ray data collected during flares as upper limits (because the X-ray and $\gamma$-ray light curves during the flare were not correlated), when they attempted to compare their model with the multiwavelength SED.  Four of the SEDs from \citet{hayash12} were also analyzed by \citet{dermer14}. The latter authors used a log-parabola electron distribution to minimize the number of free model parameters.  They also assumed equipartition between the energy densities of the magnetic field, photons, and electrons, which minimizes the jet power, and approximated the BLR contribution to EC as Ly$\alpha$, all in an attempt to ground the SED model interpretation in physics.  After all of that, the $\g$-ray data disagree with the \citet{dermer14} model above ~5 GeV, prompting the authors to suggest a second synchrotron or hadronic component as possible remedies.  

We build on each of these models in analyzing the same spectral data (2008 Aug 4 - 2009 Feb 23), but we calculate our electron distribution by solving the electron transport equation, in a more physically realistic, fully self-consistent model.  We do not assume equipartition between any energy densities, and have a more physically realistic, stratified BLR model.

\subsection{Overview}

The details of our model calculation can be found in Section \ref{model}, including the transport equation, electron distribution solution. The spectral calculation methods for each individual component are in Section \ref{emis}.   In Section \ref{paramstudy} we perform several parameter studies to illustrate the capabilities of the electron distribution solution for physical interpretation, and highlight the flexibility of the spectral calculations.  We present our comparison to 3C 279 in Section \ref{3c279}, and discuss the significance of our results in Section \ref{discuss}.

\section{Particle Transport Model}
\label{model}

The theoretical model we focus on here describes emission from a homogenous leptonic zone in a blazar jet.  We begin by developing a differential transport equation including particle injection, escape, acceleration, and energy loss terms in Section \ref{partdist}.  We solve this differential equation analytically in the Thomson regime, and numerically for the full Compton calculation, using a Runge-Kutta routine \citep[e.g.,][]{press92}, to arrive at the calculated electron distribution.  We introduce a new method for normalizing the branched steady-state solution (see Appendix \ref{ap-norm}). The electron distribution is used to calculate each spectral component in the steady-state SED in Section \ref{emis}.  Since the electron transport equation is based on first principles, and it includes radiative losses, the resulting SED is self-consistent, lending additional weight to the physical interpretation of the model parameters resulting from comparison with the observations.

\subsection{Electron Transport Equation}
\label{partdist}

The evolution of electrons in the jet blob can be described by a Fokker-Planck equation.  We use this equation to treat particle energization via diffusive shock and stochastic acceleration; energy losses from adiabatic and radiative processes; particle escape via Bohm diffusion; and the continuous injection of monoenergetic electrons.  In this subsection, all of our calculations occur in the co-moving frame, which we generally denote by primes (e.g., $N\p_e(\g\p)$), but suppress for simplicity where there are no other frames under consideration.   Our Fokker-Planck equation takes the form
\begin{align}
\frac{\partial \Ngreen}{\partial t} = & \frac{\partial^2}{\partial \gamma^2} \left( \frac{1}{2}
\frac{d \sigma^2}{d t} \Ngreen \right) - \frac{\partial}{\partial \gamma}
\left( \left< \frac{d \gamma}{d t} \right> \Ngreen \right) \nonumber \\ 
& - \frac{\Ngreen \gamma D_0}{\tau} + \dot{N}_{\rm inj} \delta(\gamma-\gamma_{\rm inj})  \ ,
\label{eq-transport}
\end{align}
where $\Ngreen(\g,t)$ is the electron number distribution and $\g \equiv E/(m_ec^2)$ is the electron Lorentz factor, with $m_e$ and $c$ denoting the electron mass and the speed of light, respectively. The quantities $\dot N_{\rm inj}$ and $\gamma_{\rm inj}$ in Equation~(\ref{eq-transport}) represent the particle injection rate and the Lorentz factor of the injected monoenergetic electrons, respectively.

We can obtain the solution to the steady-state problem by setting
\begin{equation}
\frac{\partial \Ngreen}{\partial t} = 0 \ ,
\label{eq-steady}
\end{equation}
so that $\Ngreen(\g,t)\rightarrow \Ngreen(\g)$.  We describe the individual terms below, while detailed derivations can be found in \citet{lewis16}.

The particle injection mechanism is not well understood, but we assume here that the electrons accelerated inside the blob originate as members of a low-energy thermal distribution, and that the acceleration  and emission regions are co-spatial. Since the model considered here includes a significant component of stochastic particle acceleration, represented by the second-order term in Equation~(\ref{eq-transport}), the steady-state electron distribution (the solution to Equation~(\ref{eq-steady})) will lose any ``memory'' of the energy of the injected electrons. In this scenario, the steady-state energy distribution is independent of the precise form of the energy distribution of the injected electrons. We can take advantage of this insensitivity by assuming that the injected electrons have a monoenergetic distribution, with a very low injection energy. The monoenergetic injection is represented by the $\delta$-function in Equation~(\ref{eq-transport}).

We adopt the hard-sphere approximation, so that the Fokker-Planck ``broadening coefficient'' in Equation~(\ref{eq-transport}) is given by
\begin{equation}
\frac{1}{2} \frac{d \sigma^2}{d t} = D_0 \, \g^2 \ ,
\label{eq-broadcoef}
\end{equation}
where $D_0 \propto {\rm s}^{-1}$ is the energy-diffusion coefficient \citep{park95}. We use two forms of the drift coefficient, describing the mean net acceleration rate.  They differ in their description of the Compton cooling term.  In the Thomson approximation, the drift coefficient takes the form
\begin{equation}
\left< \frac{d \gamma}{d t} \right> = D_0 \left[4\gamma+a\gamma-b_{\rm syn}\gamma^2 - \gamma^2 \sum_{j=1}^J b_{\rm C}^{(j)} \right] \ ,
\label{eq-driftThom}
\end{equation}
where $D_0$ is the hard-sphere scattering coefficient for collisions with MHD waves.  The first term on the right-hand side of Equation~(\ref{eq-driftThom}) expresses the mean energization rate due to stochastic acceleration, and the term containing $a$ accounts for first-order Fermi acceleration at the shock, as well as adiabatic cooling. Shock acceleration dominates over adiabatic losses if $a>0$; otherwise adiabatic cooling dominates.

Synchrotron cooling is expressed by the term containing $b_{\rm syn}$ in Equation~(\ref{eq-driftThom}), which is defined by
\begin{equation}
b_{\rm syn} \equiv \frac{|\dot{\gamma}_{\rm syn}|}{D_0 \gamma^2} = \frac{\sigma_{\rm T}B^2}{6 \pi m_e c D_0} \ ,
\end{equation}
where $\sigma_{\rm T} $ is the Thomson cross-section. The term containing $b_{\rm C}^{(j)}$ in Equation~(\ref{eq-driftThom}) expresses energy losses due to the scattering of external seed photons impinging on jet and colliding with the electrons in the blob. In our calculation, there are multiple sources of seed photons for Compton scattering, and each is considered independently.  Hence the Compton term is a sum over $J$ different seed photon sources, with associated dimensionless coefficients
\begin{equation}
b_{\rm C}^{(j)} \equiv \frac{|\dot{\gamma}_{\rm C}|}{D_0 \gamma^2} =\frac{4 \sigma_{\rm T} \G^2u_{\rm ph}^{(j)}}{3 m_e c D_0} \ ,
\end{equation}
where $u_{\rm ph}^{(j)}$ is the photon energy density for each seed photon source (see Section \ref{ECblr}).  

\begin{figure}
\vspace{2.2mm} 
\centering
\epsscale{1.1} 
\plotone{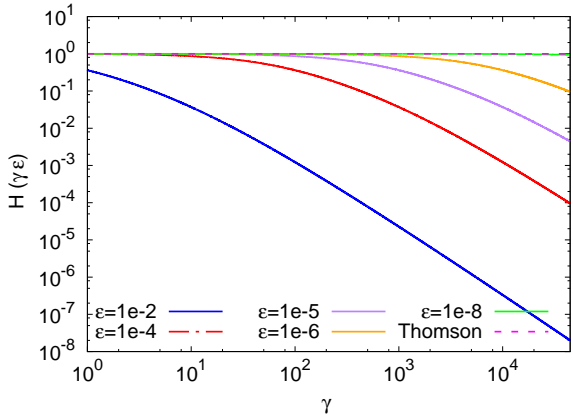}
\caption{The full Compton form from the cooling coefficient $H(\gamma\epsilon_{\rm ph})$, for a variety of dimensionless incident photon energies $\epsilon_{\rm ph}$, are compared to the Thomson approximated cooling coefficient, where $H(\gamma\epsilon_{\rm ph}) = 1$, as a function of the Lorentz factor $\gamma$.}
\label{fig-KN_coolterm}
\vspace{2.2mm}
\end{figure}

We also consider a more advanced version of the model in which we utilize the full Compton cross-section, including the KN correction. In this case, the expression for the drift coefficient can be written as
\begin{equation}
\left< \frac{d \gamma}{d t} \right> = D_0 \left[4\gamma+a\gamma-b_{\rm syn}\gamma^2 - \gamma^2 \sum_{j=1}^J b^{(j)}_{\rm C} H(\gamma \epsilon_{\rm ph}^{(j)}) \right] \ ,
\label{eq-driftKN}
\end{equation}
where $\epsilon_{\rm ph}^{(j)}$ is the incident photon energy in units of $m_ec^2$.  The function $H$ gives the energy loss rate with the full Compton cross-section included, and is defined by
\begin{equation}
H(y) \equiv \frac{9}{32}\frac{1}{y^2} G_{\rm BMS}(y) \ ,
\label{eq-Hfun}
\end{equation}
where \citep{boett97}
\begin{align}
G_{\rm BMS}&(y) = \frac{8}{3} y \frac{1+5y}{(1+4y)^2} - \frac{4y}{1+4y}  \bigg( \frac{2}{3} + \frac{1}{2y} + \frac{1}{8y^2} \bigg) \nonumber \\
& +  {\rm ln}(1+4 y)  \bigg[1+\frac{3}{y} + \frac{3}{4y^2} + \frac{{\rm ln}(1+4y)}{2y} - \frac{{\rm ln}(4y)}{y}  \bigg] \nonumber \\
& - \frac{5}{2y} - \frac{\pi^2}{6y} - 2 +\frac{1}{y} \sum_{n=1}^{\infty} \frac{(1+4y)^{-n}}{n^2} \ .
\end{align}
We note that $H(y) \to 1$ in the Thomson limit $y \ll 1$ (see Figure \ref{fig-KN_coolterm}).  Thus, Equation (\ref{eq-driftKN}) reduces to Equation (\ref{eq-driftThom}) in the Thomson regime.

Following \citet{lewis16}, our transport equation also includes a term describing the energy-dependent escape of electrons from the blob via Bohm diffusion, which can be written as
\begin{equation}
\frac{\partial \Ngreen}{\partial t}\bigg|_{\rm esc} = - \frac{\Ngreen \gamma D_0}{\tau} \ ,
\label{eq-esc}
\end{equation}
where the dimensionless escape timescale constant $\tau$ is defined in the Bohm limit as  
\begin{equation}
\tau \equiv \frac{R^{\prime 2}_{b}qBD_0}{m_ec^3} \ 
\end{equation}
\citep{lewis16}.
The final term on the right hand side of Equation (\ref{eq-transport}) is the source term, representing the injection of $\dot{N}_{\rm inj}$ electrons per second with Lorentz factor $\gamma_{\rm inj}$.  The value of $\dot{N}_{\rm inj}$ is related to the electron injection luminosity, $L_{\rm e,inj}$, by 
\begin{equation}
L_{\rm e,inj} = m_ec^2 \gamma_{\rm inj} \dot{N}_{\rm inj} \ .
\label{eq-injLum}
\end{equation}

The Compton cooling term in Equation (\ref{eq-transport}) is an addition to the drift coefficient terms previously treated in \citet{lewis16}.  While the previous equation was analytically solvable, the inclusion of the KN formalism leaves the transport equation without an analytical solution, motivating the numerical treatment employed here (see Section \ref{solnmeth}).

\subsection{Thomson Regime Solution}
\label{anasoln}

In \citet{lewis16}, we have previously solved Equation (\ref{eq-transport}) analytically in the Thomson regime ($\g\e_{\rm ph}\ll1$), i.e., if the drift coefficient is given by Equation (\ref{eq-driftThom}).  Our solution, stated in the context of this paper, can be written as
\begin{align}
\Ngreen(\g) = \frac{\dot{N}_{\rm inj} e^{b_{\rm tot}\g_{\rm inj}/2}}{b_{\rm tot}D_0 \g_{\rm inj}^{2+a/2}} \frac{\G(\sigma - \lambda +1/2)}{\G(1+ 2\sigma)} e^{-b_{\rm tot}\g/2}\g^{a/2} \nonumber \\
\times \begin{cases}
M_{\lambda,\sigma}(b_{\rm tot}\g) W_{\lambda,\sigma}(b_{\rm tot}\g_{\rm inj}), \g \le \g_{\rm inj}\\
M_{\lambda,\sigma}(b_{\rm tot}\g_{\rm inj}) W_{\lambda,\sigma}(b_{\rm tot}\g), \g \ge \g_{\rm inj}
\end{cases}
\ ,
\label{eq-anasoln}
\end{align}
 \citep{lewis16} where the total loss coefficient, $b_{\rm tot}$, is defined by
\begin{flalign}
b_{\rm tot} \equiv b_{\rm syn}+\sum_{j=1}^J  b_{\rm C}^{(j)}\ .
\end{flalign}
The indices of the Whittaker functions, $M_{\lambda,\sigma}$ and $W_{\lambda,\sigma}$, in the branching solution are given by \citep[e.g.,][]{abramowitz72}
\begin{equation}
\lambda=2 - \frac{1}{b_{\rm tot}\tau} + \frac{a}{2},\quad {\rm and } \quad \sigma=\frac{a+3}{2} \ .
\label{eq-lambda-sigma}
\end{equation}

While the form of the solution in Equation~(\ref{eq-anasoln}) is not valid once we consider KN effects, it does show that the electron distribution is dependent upon many of the same physical parameters as the emission spectra, especially the magnetic field and the photon energy densities.

\subsection{Full Compton Solution Methodology}
\label{solnmeth}

With the full Compton energy loss term included (Equation (\ref{eq-driftKN})), we can no longer solve Equation (\ref{eq-transport}) analytically.  Hence, we utilized a fourth-order Runge-Kutta finite-differencing scheme, outlined in \citet{press92}, modified to treat a second-order ODE.

We employ a grid that is uniform in terms of the logarithm of the electron Lorentz factor $\gamma$.  The minimum Lorentz factor is $\gamma_{\rm min} = 10^{0}$ in all of our calculations, and Lorentz factor of the injected electrons is set at $\gamma_{\rm inj} = 1.01$. One of our goals here is to attempt to explain the flares SEDs using a first-principles approach, without any requirement for electron pre-acceleration. Hence we choose a small value for the electron injection energy in order to model injection from the thermal electron distribution, rather than invoking any form of electron pre-acceleration, such as acceleration at shocks, which is assumed in many previous studies. The maximum Lorentz factor $\gamma_{\rm max}$ is set to accommodate the full dynamic range of the solution, and is typically between $10^{4.5}$ and $10^{7.5}$. 

Solutions to this class of transport equation follow two distinct branches for $\gamma \le \gamma_{\rm inj}$ and $\gamma \ge \gamma_{\rm inj}$.  The solution for the electron distribution, $N_e(\gamma)$, must be continuous at $\gamma_{\rm inj}$, but the derivative $\partial N_e / \partial \gamma$ is discontinuous, as discussed in detail by \citet{lewis16}.

We have developed a process for computing the electron distribution in a piecewise fashion. At the high- and low-energy boundaries, we have used the flux-free condition, i.e., the electrons should not cross the boundary at $\gamma =\gamma_{\rm min}= 1$ or $\g = \g_{\rm max}$. Once the fundamental solutions are obtained in regions above and below the injection energy, one normally utilizes the derivative-jump condition for $N_e(\gamma)$ at the injection energy $\gamma_{\rm inj}$. However, we found that the accuracy required by this method was prohibitive, due to the stiffness of the differential equation, and therefore we employed an alternative method to normalize the global solution for $N_e(\gamma)$ that is based on the requirement of a steady-state balance between electron injection and escape. This is discussed in detail in Appendix \ref{ap-norm}.  It requires much less precision than the Wronskian method, and may be useful in other steady-state applications involving stiff ordinary differential equations.
We used the analytic Thomson regime solution (Equation~(\ref{eq-anasoln})) to verify the machinery for the full Compton solution in the Thomson limit.

\subsection{Comparison of Electron Distributions}
\label{TvKNed}

\begin{figure}
\vspace{2.2mm} 
\centering
\epsscale{1.1} 
\plotone{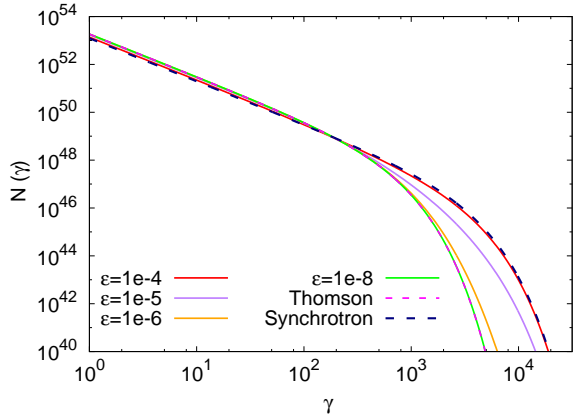}
\caption{The electron numbers distributions, $N_e(\gamma)$, obtained for varying the incident photon energy, $\epsilon_{\rm ph}$, in the EC interaction. We observe two limiting cases: where Thomson cooling dominates ($\epsilon_{\rm ph} \ll 1$) and where synchrotron cooling dominates ($\epsilon_{\rm ph} \approx 0.01$).  Between those curves, the solution for the full Compton cross-section varies, depending on the incident photon energy $\epsilon_{\rm ph}$.  Each Compton curve was produced with a single EC source located in the BLR.}
\label{fig-EvolutionEps}
\vspace{2.2mm}
\end{figure}

In Equation (\ref{eq-Hfun}),  recall that $H(y) = 1$ in the Thomson limit ($y \ll 1$), but the realization of this limit depends on both the electron Lorentz factor, $\g$, and the incident photon energy, $\e_{\rm ph}$, since $y=\gamma\epsilon_{\rm ph}$ in the full Compton calculation.  As the incident photon energy $\epsilon_{\rm ph}$ decreases, the Compton cooling term eventually approaches the Thomson limit (Figure \ref{fig-KN_coolterm}).  

It is interesting to contrast the results obtained for the electron distribution $N_e(\gamma)$ under the influence of pure synchrotron cooling, versus synchrotron plus Thomson cooling (Section \ref{anasoln}), or synchrotron plus the full Compton cooling rate (Section \ref{solnmeth}) due to a single external photon source with various values of $\e_{\rm ph}$. This comparison is carried out in Figure \ref{fig-EvolutionEps}.  For our choice of parameters (Tables \ref{table1} and \ref{table2}), for low $\e_{\rm ph}$ where the Compton solution approaches the Thomson regime, and in the Thomson regime, Compton cooling dominates ($\G^2u_{\rm ph} / [B^2/(8\pi)] \gg 1$).  However, as $\e_{\rm ph}$ increases, the Compton cooling at high $\g$ becomes less efficient (Figure \ref{fig-KN_coolterm}), and therefore synchrotron losses begin to dominate.  For $\e_{\rm ph}\le10^{-4}$, the solution is indistinguishable from the synchrotron-only case.

\begin{deluxetable}{lllr}
\tabletypesize{0.5\scriptsize}
\tablecaption{Physical \& Assumed Parameters for 3C 279}
\tablewidth{0pt}
\tablehead{
\colhead{Variable (Unit)}
& \colhead{Symbol}
& \colhead{Value}
}

\startdata
Energy of Ly$\alpha$
&$\epsilon_{\rm Ly\alpha}$
&$2.0 \times 10^{-5}$
\\
Luminosity Distance (cm)
&$d_{\rm L}$
&$9.61\times10^{27}$
\\
Redshift
&$z$
&$0.536$
\\
\hline
\\
Dust Efficiency
&$\xi$
&$0.1$
\\
Particle Injection Energy
&$\gamma_{\rm inj}$
&$1.01$
\enddata
\label{table1}
\end{deluxetable}
\vspace{2.2mm}

\begin{deluxetable}{lllr}
\tabletypesize{0.5\scriptsize}
\tablecaption{Base Parameters for Parameter Study}
\tablewidth{0pt}
\tablehead{
\colhead{Variable (Unit)}
& \colhead{Symbol}
& \colhead{Base Value}
}

\startdata
Variability Timescale (s)
&$t_{\rm var}$
&$2.0 \times 10^{4}$
\\
Magnetic Field (G)
&$B$
&$1.0$
\\
Doppler Factor
&$\delta_{\rm D}$
&$25.0$
\\
Energy Density of Ly$\alpha$ (erg cm$^{-3}$)
&$u_{\rm Ly\alpha}$
&$2.5\times 10^{-4}$
\\
Dust Temperature (K)
&$T_{\rm dust}$
&$1400$
\\
Disk Luminosity (erg s$^{-1}$)
&$L_{\rm disk}$
&$1.0 \times 10^{46}$
\\
Second-Order Fermi Coefficient
&$D_0$
&$2.0\times 10^{-6}$
\\
First-Order Fermi Coefficient
&$a$
&$-3.8$
\\
Energy Injection Rate
&$L_{\rm inj}$
&$8.3 \times 10^{29}$
\enddata
\label{table2}
\end{deluxetable}

\section{Calculation of Emission Spectrum}
\label{emis}

\begin{figure*}
\vspace{2.2mm} 
\centering
\includegraphics[width=\textwidth, trim={0 0 0.4cm 4.1cm},clip] {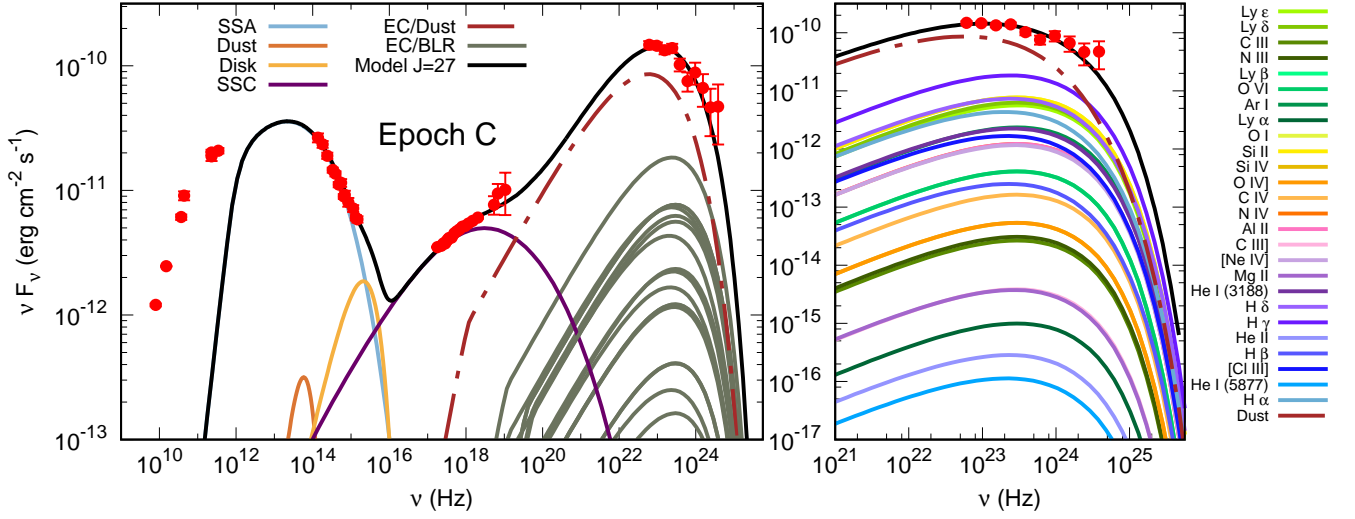}
\caption{The $\nu F_{\nu}$ SED for 3C 279, showing the individual spectral components.  The left panel, from left to right, depicts the spectra for synchrotron radiation (including self-absorption), disk emission, SSC, and EC for different seed photon sources. The right panel shows all of the individual EC components in detail, from dust and each individual BLR emission line. The emission model examined here reproduces ``epoch C'' of 3C 279, with parameters given in Table \ref{table3}.}
\label{fig-SEDwBLR}
\vspace{2.2mm}
\end{figure*}

We include emission from three distinct regions in our SED modeling:  a thermal accretion disk, a dust torus, and the jet blob.  Since we are only interested in simulating the primary emission component, and the radio is likely formed from a larger jet component, for our purposes, the radio data are upper limits.  The computation of the various spectral components is discussed below.

\subsection{Disk Emission}
\label{disk}

The disk spectrum is approximated using the \citet{shakura73} formalism, which yields
\begin{equation}
\label{eq-disklum}
\e L^{\rm disk}(\e) = 1.12\ \left( \frac{ \e}{ \e_{\rm max}} \right)^{4/3} {\rm e}^{-\e / \e_{\rm max}} \ ,
\end{equation}
\citep[e.g.,][]{dermer14} where $\e$ is the stationary frame dimensionless energy and $m_e c^2 \e_{\rm max} = 10$ eV.  The observed $\nu F_\nu$ spectrum from the disk will then be
\begin{flalign}
f^{\rm disk}_{\e_{\rm obs}} = \frac{ (1+z) \e_{\rm obs} L^{\rm disk}( (1+z) \e_{\rm obs}) }{ 4\pi d_L^2 } \ ,
\end{flalign}
where the factors of $(1+z)$ take into account the cosmological redshift.

\subsection{Dust Torus}

The spectrum radiated by the dust torus is assumed to be an infrared blackbody, so that
\begin{flalign}
\e L^{\rm dust}(\e) = \frac{15 L^{\rm dust}}{\pi^4} \frac{(\e/\Theta)^4}{\exp(\e/\Theta) - 1} \ ,
\end{flalign}
where the dimensionless dust temperature $\Theta = k_{\rm B}T_{\rm dust}/(m_ec^2)$, and $k_{\rm B}$ denotes the Boltzmann constant. Hence the observed dust torus spectrum is given by
\begin{flalign}
f^{\rm dust}_{\e_{\rm obs}} = \frac{ (1+z) \e_{\rm obs} L^{\rm dust}((1+z) \e_{\rm obs}) }{ 4\pi d_L^2 }\ 
\end{flalign}
\citep[e.g.,][]{dermer14}.

\subsection{Emission from Jet Blob}

In addition to the disk and torus components, the observed blazar SED also includes emission from the nonthermal electrons in the plasma blob, which is propagating inside the relativistic jet. These components are due to synchrotron, SSC, and external Compton-scattered emission.  Once we have solved the steady state transport equation (Equation (\ref{eq-transport})) to determine the electron distribution in the co-moving frame of the blob, $N\p_e(\gp)$, we can use that solution to calculate the jet emission components due to synchrotron, SSC, EC/Dust, and EC/BLR.  Our goal is to use our steady-state model to interpret the SED of 3C 279 observed at different epochs.  Each epoch will have a different set of physical parameters, and a different electron distribution.

In general, quantities calculated in the co-moving (blob) frame are primed, and those in the stationary (BH) frame or observer (Earth) frame are unprimed.  Quantities in the observer frame will be affected by the relativistic Doppler effect \citep[e.g.,][]{r&l79} as well as the cosmological redshift.  In general, the dimensionless energy in the observer frame is related to that in the co-moving frame via
\begin{equation}
\epsilon_{\rm obs} = \frac{\delta_{\rm D}}{1+z}\ \ep \ .
\label{eq-EpsPrime}
\end{equation}

\subsubsection{Synchrotron Emission}

In the observer frame, the $\nu F_{\nu}$ synchrotron emission spectrum is given by
\begin{flalign}
f_{\epsilon_{\rm obs}}^{\rm syn} = 
\frac{\sqrt{3} \e' \delta_{\rm D}^4 e^3 B}{4\pi h d_{\rm L}^2} 
\int^\infty_1 d\gp\ N\p_e(\gp)\ R(x)\ ,
\label{fsy}
\end{flalign}
where 
\begin{flalign}
x = \frac{4\pi \epsilon' m_e^2 c^3}{3eBh\g^{\prime 2}}\ ,
\label{eq-x}
\end{flalign}
and
\begin{flalign}
R(x) \equiv \frac{x}{2}\int_0^\pi d\theta\ \sin\theta\ 
\int^\infty_{x/\sin\theta} dw\ K_{5/3}(w)\ ,
\label{eq-R}
\end{flalign}
\citep{crusius86}. Here, $K_{5/3}(w)$ denotes the modified Bessel function of order 5/3.  Numerically, we use the approximate expression for $R(x)$ from \citet{finke08_SSC}; other useful approximations are given by \citet{zirak07} and \citet{joshi11}.

We also consider the self-absorption of the synchrotron radiation, which depletes the low-energy portion of the synchrotron spectrum.  The self-absorbed spectrum in the observer frame, denoted by $f_{\epsilon_{\rm obs}}^{\rm syn,abs}$, is computed in the slab approximation using
\begin{equation}
f_{\epsilon_{\rm obs}}^{\rm syn,abs} = \frac{1- {\rm e}^{-\tau^{\rm SSA}}}{\tau^{\rm SSA}}
f_{\epsilon_{\rm obs}}^{\rm syn} \ ,
\end{equation}
\citep[e.g.,][]{d&m09} where $f_{\epsilon_{\rm obs}}^{\rm syn}$ is computed using Equation~(\ref{fsy}), and where the synchrotron self-absorption optical depth, $\tau^{\rm SSA}$, is given by 
\begin{equation}
\tau_{\rm SSA} = \frac{- r\p_b c^2}{8\pi \nu m_e c^2} \int_1^{\infty}P_{\nu}(\g) \g^2 \, \frac{d}{d\g} \! \left[ \frac{n(\g)}{\g^2}\right]  d\g  \ ,
\end{equation}
\citep[e.g.][]{boett97} where
\begin{equation}
P_{\nu}(\g) = \frac{\sqrt{3}e^3B}{m_ec^2R(x)} \ ,
\end{equation}
\citep[e.g.][]{r&l79} and $R(x)$ is given in Equation (\ref{eq-R}).

\subsubsection{Synchrotron Self-Compton}

Synchrotron radiation also serves as a source of seed photons for the SSC process, with corresponding spectrum
\begin{flalign}
f_{\e_s}^{\rm SSC} & = \frac{9}{16} \frac{(1+z)^2 \sT \e_s^{\prime 2}}
{\pi \dD^2 c^2t_{v}^2 } 
 \int^\infty_0\ d\ep_*\ 
\frac{f_{\e_*}^{\rm syn}}{\e_*^{\prime 3}}\ 
\nonumber \\ & \times
\int^{\infty}_{\gp_{1}}\ d\gp\ 
\frac{N\p_e(\gp)}{\g^{\prime2}} F_{\rm C}\left(4 \g\p \ep_{*}, \frac{\e}{\g\p} \right) \ ,
\label{fSSC}
\end{flalign}
\citep[e.g.,][]{finke08_SSC} where $\ep_s= \e_s(1+z)/\dD$, $\ep_*=\e_*(1+z)/\dD$, and
\begin{flalign}
\gp_{1} = \frac{1}{2}\epsilon\p_s \left( 1+ \sqrt{1+ \frac{1}{\epsilon\p \epsilon\p_s}} \right) \ .
\end{flalign}
The function $F_{\rm C}$ in Equation~(\ref{fSSC}) is defined by
\begin{equation}
F_{\rm C}(p,q) = 2w {\rm ln}(w) + (1 + 2w)(1 - w) + \frac{(pw)^2}{2} \frac{1 -w}{1 + pw} \ ,
\label{eq-FCqGam}
\end{equation}
\citep{jones68,blumen70} where
\begin{flalign}
w = \frac{q}{p (1-q)} \ .
\end{flalign}

\subsubsection{Compton-Scattered External Radiation}

The external-Compton SED is calculated by assuming that each external photon source (whether BLR line or dust) is monochromatic and isotropic and homogeneous in the stationary frame, so that
\begin{flalign}
\label{eq-ECflux}
f_{\e_s}^{\rm EC} & = \frac{3}{4} \frac{c\sT \e_s^2}{4\pi d_L^2}\frac{u_*}{\e_*^2} \dD^3 
\nonumber \\ & \times
\int_{\g_{1}}^{\g_{\max}} d\g
\frac{N\p_e(\g/\dD)}{\g^2}F_{\rm C} \left(4 \g \e_*, \frac{\e_s}{\g} \right) \ ,
\end{flalign}
where
\begin{flalign}
\g_{1} = \frac{1}{2}\e_s \left( 1+ \sqrt{1+ \frac{1}{\e \e_s}} \right) \ 
\end{flalign}
\citep[e.g.,][]{georgan01,dermer09}.  

We apply the external Compton formalism to radiation from the disk, the dust torus and the BLR.  In this type of calculation, \citet{dermer14} assumed that Ly$\alpha$ will be the most prominent emission line in the BLR, and therefore that all other emission lines can be neglected.  We studied this assumption in detail by implementing one version of our model in which Ly$\alpha$ was the only emission from the BLR to undergo EC, in addition to dust ($J=2$). We also implemented another version in which 25 additional lines from the \citet{vanden01} template were included as sources of BLR seed photons ($J=27$).  This latter model comprises dust seed photons plus the 26 lines for which \citet{finke16} provided radius, luminosity, and energy density relationships based on SDSS composite quasar spectra \citep{vanden01}. By adopting the formalism of \citet{finke16}, we are able to avoid increasing the number of model free parameters when including additional BLR constituents. 

Each source of EC seed photons is used to generate a corresponding cooling term in the electron transport equation (Equation~(\ref{eq-transport})), and therefore the associated components of the SED are computed self-consistently in our model. We provide further details on the calculation method for each EC spectral component below.

\subsubsection{External Compton of Disk Photons}
\label{ECdisk}

The disk emits photons in the optical/UV due to thermal radiation described in Section \ref{disk}.  In principle, any photons incident on the blob will experience Compton scattering by the blob electrons, and we estimate that effect for the disk photons by calculating the resulting $\nu F_{\nu}$ flux using
\begin{equation}
f_{\e}^{\rm EC/Disk} = \frac{r_e^2 \e_s^2 L_{\rm disk} \delta_{\rm D}^3}{16\pi r^2_{\rm blob} d^2_{\rm L}\e^2_{\rm disk}} \int_{\bar{\g}_{\rm low}}^{\infty} d\g \frac{N\p_e(\g/\delta_{\rm D})}{\g^2} \, \bar{\Xi}
\label{ECdiskeq}
\end{equation}
\citep[e.g.][]{dermer09}, where we approximate the disk spectrum as monochromatic with an energy $m_ec^2\e_{\rm disk} \approx 10$ eV.  The function $\bar{\Xi}$ in the integrand is defined by
\begin{equation}
\bar{\Xi} \equiv 1-y + \frac{1}{1-y} - \frac{2y}{\bar{\e} (1-y)} + \left[ \frac{y}{\bar{\e} (1-y)} \right]^2 \ ,
\end{equation}
where $y = \e_s/\g$ and $\bar{\e} = \g \e_{\rm disk} (1-\mu_s)$.  The lower bound of the integration in
Equation~(\ref{ECdiskeq}),
\begin{equation}
\bar{\g}_{\rm low} = \frac{\e_s}{2}\left[ 1 + \sqrt{ 1+ \frac{2}{\e_{\rm disk} \e_s (1-\mu_s)}} \,\, \right] \ ,
\end{equation}
where $\mu_s=\cos\theta$.  We assume $\theta = 1/\Gamma$ so that $\Gamma = \delta_{\rm D}$.

We implement this calculation, but find that the contribution of external Compton scattering of disk photons is negligible in comparison to the rest of the SED components.  Therefore, we neglect it in the remainder of the analysis. 

\subsubsection{External Compton of Dust Torus Photons}
\label{ECdust}

The dust torus emits an infrared distribution of blackbody photons, powered by the absorption of disk radiation.  For the purpose of Compton scattering, we can approximate the dust emission spectrum as a monochromatic line with energy given by the peak of the blackbody. The corresponding dimensionless line energy for dust temperature $T_{\rm dust}$ is
\begin{equation}
\epsilon_{\rm dust} = 5 \times 10^{-7} \bigg( \frac{T_{\rm dust}}{1000\ {\rm K}} \bigg) \ ,
\end{equation}
and the associated energy density is given by
\begin{equation}
u_{\rm dust} = 2.2 \times 10^{-5}  \bigg(\frac{\xi}{0.1} \bigg) \bigg( \frac{T_{\rm dust}}{1000\ {\rm K}} \bigg)^{5.2}  ~ {\rm erg ~ cm^{-3}} \ 
\end{equation}
\citep{nenkova08,sikora09}, where $\xi \equiv L^{\rm dust}/L^{\rm disk} \le 1$ is the efficiency with which the dust reprocesses disk photons.  In our calculations, we set $\xi = 0.1$, which is an appropriate covering factor for the torus \citep{sikora09}.  The dust temperature $T_{\rm dust}$ is allowed to vary in the fitting process, but had to be less than the dust sublimation temperature, and therefore $T_{\rm dust} \la 2000$~K.

\subsubsection{External Compton of BLR Emission Line Photons}
\label{ECblr}

The spectrum of the Compton-scattered BLR emission for each line depends crucially on the radius at which the line photons are generated in the BLR ($r_{\rm line}$), and the location of the jet blob relative to the line source.  \citet{finke16} developed a simple way to estimate the location and emission properties of the various lines found in the composite SDSS quasar template \citep{vanden01} by using empirical relations derived from reverberation mapping.  From Equation (\ref{eq-disklum}) we can find the disk luminosity for photon energy $\e$ corresponding to wavelength 5100 \AA, which we denote as $L_{5100 \angstrom}$.  Reverberation mapping indicates that the distance of the H$\beta$ source from the BH is then given by
\begin{equation}
r_{\rm H\beta} = 10^{16.94} \bigg(\frac{L_{5100\angstrom}}{10^{44}\ \erg\ \s^{-1} }\bigg)^{0.533} \qquad {\rm cm} \ ,
\label{eq-RHbet}
\end{equation}
\citep{bentz13} neglecting uncertainties.  The distance from the BH of the peak emission for a particular emission line, $r_{\rm line}$, is related to the distance of the H$\beta$ line source from the BH by 
\begin{equation}
r_{\rm line} = r_{\rm H\beta} \, \rho_{\rm line}  \ ,
\end{equation}
where the quantity $\rho_{\rm line}$ is provided in Table 5 from \citet{finke16} for all of the lines in the composite SDSS quasar spectrum of \citet{vanden01}.  This
table is predicated on the assumption that the ratios $\rho_{\rm line}$ are the same for all quasars.

Next, we note that the specific luminosity for the H$\beta$ line is given by
\begin{equation}
L_{\rm H\beta} = 1.425\times10^{42}\, {\rm erg \, s^{-1} } \left( \frac{L_{5100 \angstrom}}{10^{44} \, {\rm erg \, s^{-1} }}\right)^{1/0.8826} \ 
\end{equation}
\citep{greene05}.  The luminosity of a particular emission line, $L_{\rm line}$, is related to the luminosity of the H$\beta$ line by 
\begin{equation}
L_{\rm line} = L_{\rm H\beta} \, \ell_{\rm line} \ ,
\end{equation}
where the quantities $\ell_{\rm line}$ are tabulated in Table~5 from \citet{finke16}.

\begin{figure*}
\vspace{2.2mm}
\centering
\epsscale{1.0} 
\plotone{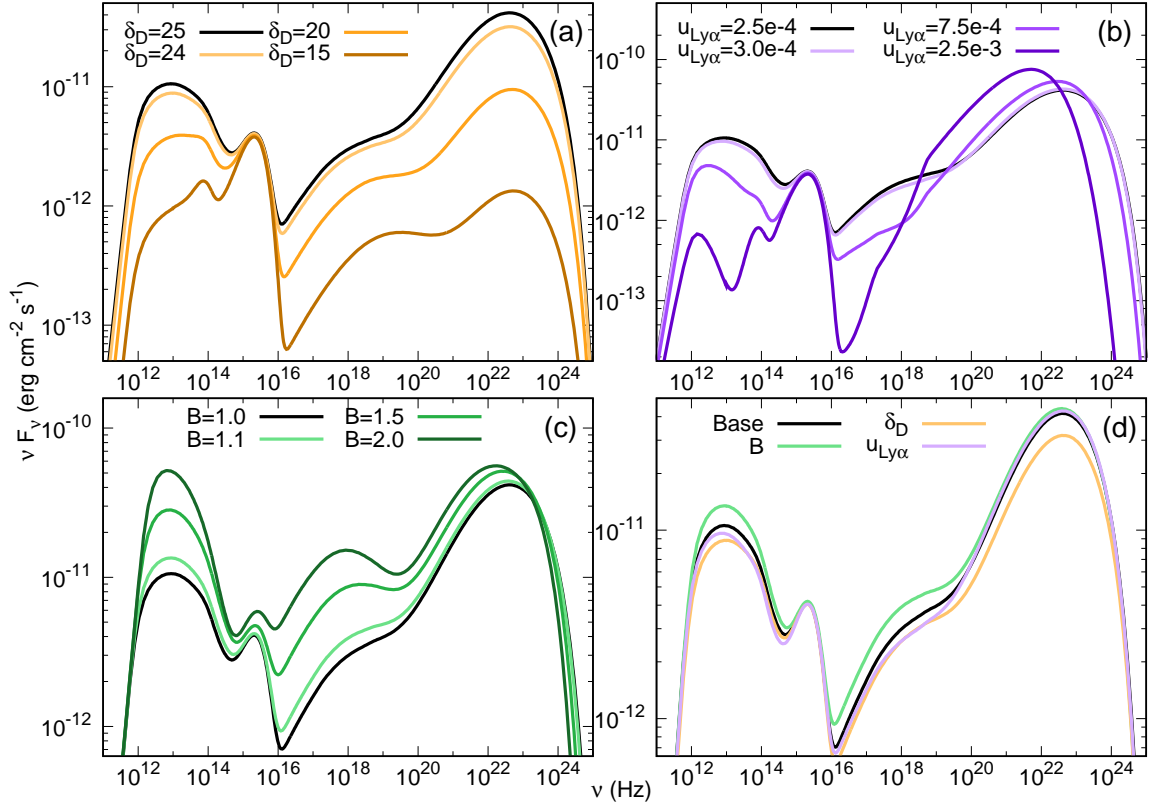}
\caption{The black line in each panel represents the SED resulting from the set of base parameters listed in Table~\ref{table2}.  Panel (a) shows the effect on the SED when the Doppler/bulk Lorentz factor $\delta_{\rm D}$ is the only parameter varied. Panel (b) shows the spectral variation when only the energy density of Ly$\alpha$ is varied. Panel (c) depicts the SED shift when the magnetic field $B$ is varied.   Panel (d) compiles the smallest changes from each of the other frames in comparison to the base set of parameters. }
\label{fig-ParamStdyPhys}
\vspace{2.2mm}
\end{figure*}

The energy density of a particular emission line can be approximated as
\begin{flalign}
u_{\rm line} = \frac{u_{\rm line,0}}{1 + (r_{\rm blob}/r_{\rm line})^\beta} \ ,
\label{eq-uline}
\end{flalign}
where
\begin{flalign}
u_{\rm line,0} = \frac{ L_{\rm line} }{4\pi c r_{\rm line}^2 }
\end{flalign}
\citep[e.g.,][]{hayash12,dermer14}.  If $ r_{\rm blob} \ll r_{\rm line}$, then $u_{\rm line} \rightarrow u_{\rm line,0}$.  Detailed geometric calculations by \citet{finke16} indicate that $\beta\approx 7.7$, which we use in all our calculations. 

We use the method above to calculate the energy densities $u_{\rm ph}^{(j)}$ for all of the BLR seed photons and also for the dust seed photons.  We use the energy densities to compute $b_{\rm C}$, the coefficient of the Compton cooling term, and to calculate the observed $\nu F_\nu$ spectrum using Equation (\ref{eq-ECflux}).  An example of our full SED calculation with all our spectral components included is provided in Figure \ref{fig-SEDwBLR}, which is further discussed in Section~\ref{3c279}.

\begin{figure*}
\vspace{2.2mm} 
\centering
\epsscale{1.0} 
\plotone{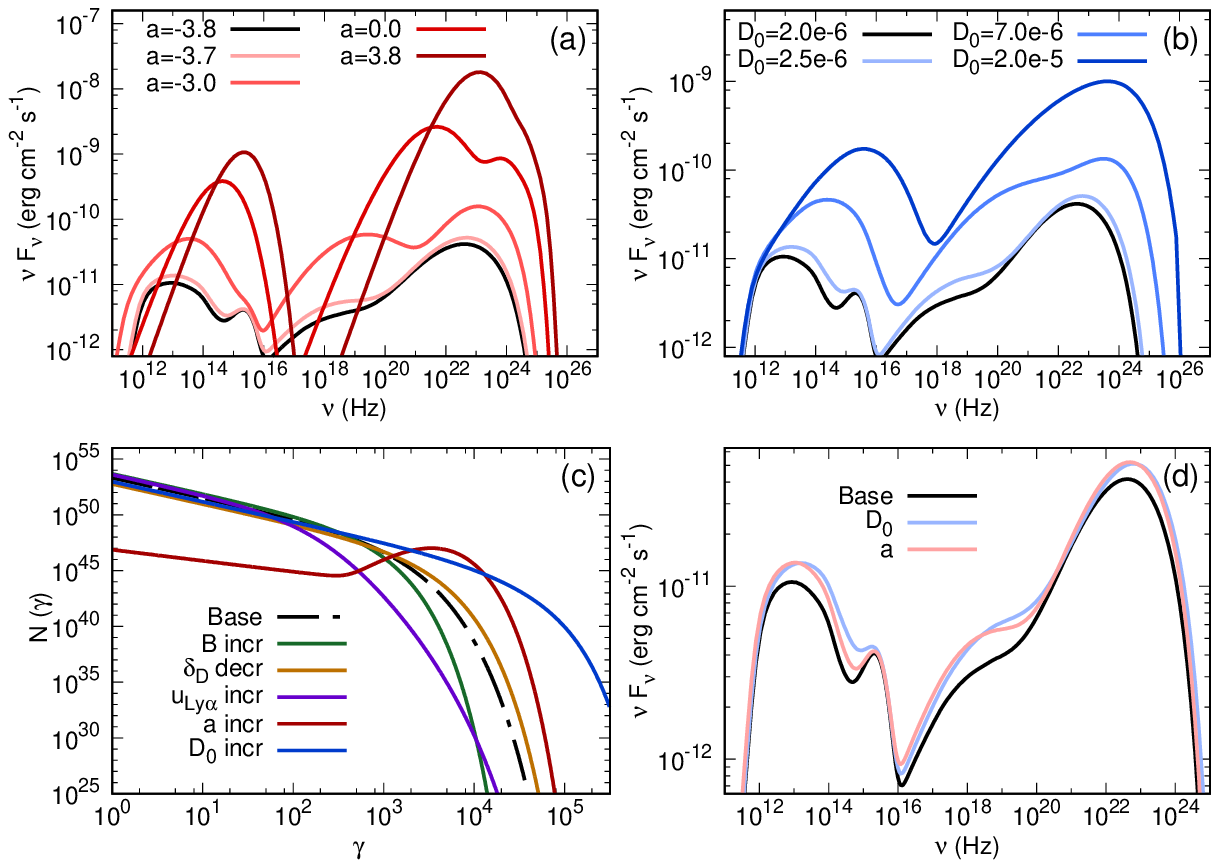}
\caption{The black line in each panel represents the SED resulting from the set of base parameters listed in Table~\ref{table2}. In panel (a), we vary only the first-order Fermi acceleration coefficient, $a$, to produce the plotted SEDs. In panel (b), we plot the SEDs resulting from variation of the second-order Fermi acceleration coefficient, $D_0$.  In panel (c), we compile all of the electron distributions resulting from the largest parameter variations considered in both Figure \ref{fig-ParamStdyPhys} and Figure \ref{fig-ParamStdyMod}. In panel (d), we compile the SEDs resulting from the smallest parameter variations of $a$ and $D_0$.}
\label{fig-ParamStdyMod}
\vspace{2.2mm}
\end{figure*}

\subsection{Jet Powers}
\label{jetpower}

We compute the stationary frame jet powers resulting from our models to verify that there is enough energy available in the accretion flow plus the rotating supermassive black hole to power the jet, and evaluate whether or not it is in equipartition. The Poynting flux (magnetic field) jet power is computed using
\begin{flalign}
P_B = 2\pi R^{\prime 2}_b\beta c\G^2 u_B \ ,
\end{flalign}
and the electron jet power is likewise given by
\begin{flalign}
P_e = 2\pi R^{\prime 2}_b\beta c\G^2 u_e \ ,
\end{flalign}
where
\begin{flalign}
u_e = \frac{m_e c^2}{V\p_b} \int d\gp\ \gp\ N\p_e(\gp) \ ,
\end{flalign}
and $V\p_b=4\pi R^{\prime 3}_b/3$. The jet powers we obtain are reported in Table \ref{table4}.

\section{Parameter Study}
\label{paramstudy}

We demonstrate the effect of the variation of several of the model free parameters on a sample SED in Figures \ref{fig-ParamStdyPhys} and \ref{fig-ParamStdyMod} by incrementally changing each parameter, independently, over a physically appropriate range. In our parameter study, the quantities listed in Table \ref{table1} are always held fixed at the indicated values, and the quantities listed in Table \ref{table2} are allowed to vary.

The parameters for the baseline model are indicated in Table \ref{table2}, and all quantities are varied around these values. Our baseline parameter values are similar, but not identical to the parameters resulting from our model comparison to quiescent data for 3C 279, which is discussed in Section~\ref{3c279}. The entire parameter study is conducted using the full Compton cross-section, but without the stratified BLR; hence we are assuming only dust and Ly$\alpha$ EC sources ($J=2$).

In all of the panels in Figures \ref{fig-ParamStdyPhys} and \ref{fig-ParamStdyMod}, the ``base value'' solution is plotted in black for comparison, and the deepening shades of color indicate stronger departures from the baseline model. Figures \ref{fig-ParamStdyPhys}d and \ref{fig-ParamStdyMod}d present a summary of the SEDs resulting from small parameter variations, and Figure \ref{fig-ParamStdyMod}c compares the electron distributions.

In Figure \ref{fig-ParamStdyPhys}, we examine the result of varying a few of the more physically explicit free parameters in our model.  In Figure \ref{fig-ParamStdyPhys}a, we vary the Doppler factor from $\delta_{\rm D} =15$ to 25, which are typical values for 3C 279.  As $\delta_{\rm D}$ decreases, the overall magnitude of the SED decreases except in the optical/UV range, where the disk emission dominates.  In the near IR, the dust torus emission begins to dominate for low values of $\dD$.  More interesting is that as $\dD$ decreases, the ratio of EC ($\gamma$-ray) to synchrotron (radio/IR) and SSC (X-ray) decreases.  This is a well-known effect, that stems from the fact that the beaming patterns for synchrotron and SSC are different from EC \citep{dermer95,georgan01}.  Additionally, as $\dD$ decreases, the synchrotron and SSC components shift to higher energies to a greater extent than the EC, effectively compressing the frequency range of the SED.

In Figure \ref{fig-ParamStdyPhys}b, we vary the energy density of the Ly$\alpha$ emission line, $u_{\rm L\alpha}$, which is the only source of EC seed photons from the BLR considered in the parameter study.  We increase the energy density, by an order of magnitude, from a base value of $u_{\rm Ly\alpha} = 2.5 \times 10^{-4}$ erg cm$^{-3}$, which is in the range of typical values for the quiescent fits.  As $u_{\rm Ly\alpha}$ increases, the ratio of EC to synchrotron emission increases, and the SSC becomes negligible.  As $u_{\rm Ly\alpha}$ increases, the EC peak shifts to lower frequencies.  The peaks of synchrotron and SSC emission shift to lower frequencies more rapidly, causing the SED to expand in frequency space.  As the synchrotron curve moves to lower energies, the entire peak becomes overwhelmed by the self-absorption cutoff in the radio regime. 

In Figure \ref{fig-ParamStdyPhys}c, we increase the magnetic field from $B = 1\,$G to $B = 2\,$G.  While the SED increases in magnitude overall, the ratio of EC to synchrotron and SSC decreases.  Additionally, the peak of the synchrotron component remains in approximately the same location, while the SSC and EC curves move to lower frequencies for higher magnetic field values.

Figure \ref{fig-ParamStdyPhys}d compares the baseline SED with those resulting from the smallest variations in the parameters $\delta_{\rm D}$, $u_{\rm L\alpha}$, and $B$ that we consider here. These results may demonstrate how changing multiple parameters at once might affect the overall SED.  For example, if we want to increase the ratio of synchrotron to EC without changing SSC, we might try to balance an increase of $B$ with a decrease of $\delta_{\rm D}$. 

\begin{deluxetable*}{lccccccccr}
\tabletypesize{\scriptsize}
\tablecaption{Free Model Parameters}
\tablewidth{0pt}
\tablehead{
\multirow{2}{*}{Parameter~(Unit)}
&\multicolumn{8}{c}{Epoch (Model)}\\
\colhead{\quad}
& \colhead{A ($J=2$)}
& \colhead{A ($J=27$)}
& \colhead{B ($J=2$)}
& \colhead{B ($J=27$)}
& \colhead{C ($J=2$)}
& \colhead{C ($J=27$)}
& \colhead{D ($J=2$)}
& \colhead{D ($J=27$)}
}
\startdata
\\
$t_{v}$ (s)
&$1.0 \times 10^{4}$
&$9.3 \times 10^{3}$
&$1.5 \times 10^{4}$
&$4.8 \times 10^{3}$
&$1.0 \times 10^{3}$
&$1.8 \times 10^{3}$
&$1.8 \times 10^{5}$
&$1.8 \times 10^{5}$
\\
$B$ (G)
&$1.24$
&$1.25$
&$1.22$
&$1.65$
&$1.30$
&$1.37$
&$0.54$
&$0.56$
\\
$\delta_{\rm D}$
&$30$
&$30$
&$39$
&$55$
&$90$
&$70$
&$20$
&$20$
\\
$r_{\rm blob}$ (cm)
&$1.64 \times 10^{17}$
&$5.55 \times 10^{17}$
&$1.84 \times 10^{17}$
&$4.69 \times 10^{17}$
&$1.30 \times 10^{17}$
&$7.00 \times 10^{17}$
&$1.83 \times 10^{17}$
&$5.99 \times 10^{17}$
\\
$T_{\rm dust}$ (K)
&$1410$
&$1450$
&$1420$
&$1290$
&$850$
&$1100$
&$1500$
&$1480$
\\
$L_{\rm disk}$ (erg s$^{-1}$)
&$7.5 \times 10^{45}$
&$9.0 \times 10^{45}$
&$8.0 \times 10^{45}$
&$5.0 \times 10^{45}$
&$3.0 \times 10^{45}$
&$5.0 \times 10^{45}$
&$6.5 \times 10^{45}$
&$6.5 \times 10^{45}$
\\
$D_0$ (s$^{-1}$)
&$3.2 \times 10^{-6}$
&$3.8 \times 10^{-6}$
&$4.0 \times 10^{-6}$
&$4.5 \times 10^{-6}$
&$2.0 \times 10^{-6}$
&$3.3 \times 10^{-6}$
&$2.2 \times 10^{-6}$
&$2.3 \times 10^{-6}$
\\
$a$
&$-3.8$
&$-3.8$
&$-3.6$
&$-3.5$
&$-4.1$
&$-3.9$
&$-3.7$
&$-3.7$
\\
$L_{\rm inj}$ (erg s$^{-1}$)
&$7.36 \times 10^{29}$
&$7.28 \times 10^{29}$
&$1.74 \times 10^{29}$
&$1.12 \times 10^{29}$
&$6.80 \times 10^{29}$
&$4.13 \times 10^{29}$
&$3.31 \times 10^{29}$
&$3.18 \times 10^{29}$

\enddata
\label{table3}
\end{deluxetable*}


In Figure \ref{fig-ParamStdyMod}a, we explore a range of values for the first-order Fermi coefficient, $a$, which represents the combined effect of shock acceleration and adiabatic losses.  When $a > 0$, shock acceleration dominates; otherwise, adiabatic cooling dominates.  We observe that as $a \rightarrow 0$ from the negative side, the SED broadens, and the ratios of EC to synchrotron and SSC decrease.  The ratio of EC to SSC decreases more than that of the synchrotron, and  SSC shifts to higher frequencies.  As shock acceleration begins to dominate over adiabatic expansion ($a$ becomes positive), the entire SED shifts to higher frequencies, with SSC moving more than either synchrotron or EC, such that SSC is the dominant component in the $\gamma$-ray regime when shock acceleration is dominant.  Additionally, we note that the peak of the spectral features narrows significantly as the effect of shock acceleration is amplified.

Figure \ref{fig-ParamStdyMod}b depicts the effect of an order of magnitude increase in the MHD wave-electron scattering (or second order Fermi acceleration) coefficient, $D_0$.  We note that as $D_0$ increases, the SED broadens overall, while the peak of each spectral component also becomes broader. The peak of the synchrotron and SSC components move further toward higher frequencies than the EC, such that EC and SSC combine into a single broad $\gamma$-ray peak when $D_0 = 2.0 \times 10^{-5}$.  Note that the synchrotron curve is more symmetric when the peak is further from the SSA cutoff.  Additionally, the nonthermal spectral bumps are broader for increased stochastic acceleration in Figure \ref{fig-ParamStdyMod}b, while they are narrower for increased shock acceleration in Figure \ref{fig-ParamStdyMod}a.  This is more apparent in the synchrotron curve of each panel, since the SSC and EC blend in both scenarios, but it is true in general.

In Figure \ref{fig-ParamStdyMod}c, we plot the electron distribution for the base set of parameters, and also for each of the most extreme changes from that base set that were used to create the SEDs plotted in Figures \ref{fig-ParamStdyPhys} and \ref{fig-ParamStdyMod}. Increasing $D_0$ extends the electron distribution to higher energies.  Increasing $a$ changes the electron distribution shape to produce a pile-up of high energy electrons, while depleting the number at lower energies.  Notably, this bump is due to the efficient acceleration of electrons in the shock, rather than the decreased efficiency of nonthermal cooling in the KN regime, as one might expect \citep[e.g.,][]{dermer02_KN,moderski05}.  The Thomson version of the model yields the same bump, albeit peaking at lower energies.  Increasing $u_{\rm Ly\alpha}$ broadens the knee feature, as full Compton cooling begins to separate from synchrotron cooling in the electron energy space. By comparison, the increase in $B$ and the decrease in $\delta_{\rm D}$ provided relatively minor changes to the overall electron distribution. 

In Figure \ref{fig-ParamStdyMod}d, we compare the base-model SED, with those obtained when small changes in $D_0$ or $a$ are invoked (cf. Figure \ref{fig-ParamStdyPhys}d).  The plots suggest how changing multiple acceleration parameters at once might affect the overall SED, especially in the synchrotron component, since the high-energy slope is sensitive to these parameter variations.  For example, to broaden the synchrotron bump, we might increase $D_0$ while decreasing $a$.

\section{Application to 3C 279 in 2008-2009}
\label{3c279}

\begin{figure}
\vspace{2.2mm} 
\centering
\epsscale{1.1} 
\plotone{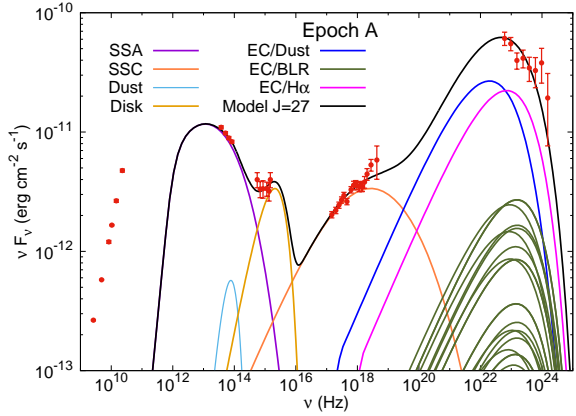}
\caption{SED data in red (from Epoch A of \citet{hayash12}) compared with our steady-state model (black curve), computed using the full BLR implementation ($J=27$).  Note that all of the EC/BLR components are united in color as a single process, except the most prominent, H$\alpha$, which we highlight as indicated.}
\label{fig-A_SED}
\vspace{2.2mm}
\end{figure}

\begin{figure}
\vspace{2.2mm} 
\centering
\epsscale{1.1} 
\plotone{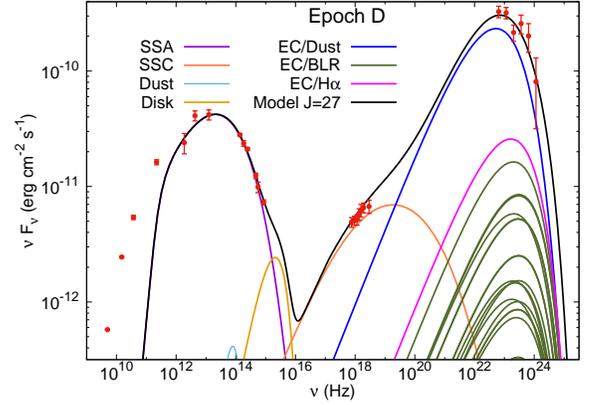}
\caption{Same as Figure~\ref{fig-A_SED}, except the SED data in red are from Epoch D of \citet{hayash12}.}
\label{fig-B_SED}
\label{fig-D_SED}
\vspace{2.2mm}
\end{figure}

We demonstrate the new features of our model by using it to qualitatively fit and interpret epochs A--D of the 3C 279 spectral data reported by \citet{hayash12}.   We utilize our most realistic model, which is the stratified BLR model with $J=27$.  The results for epoch C are depicted in Figure \ref{fig-SEDwBLR}, and our results for epochs A, B, and D are reported in Figures \ref{fig-A_SED}, \ref{fig-B_SED}, and \ref{fig-D_SED}, respectively.  The values of the model free parameters for each case are listed in Table \ref{table3}, and associated calculated parameters are listed in Table \ref{table4}. We discuss some of our detailed findings for each epoch below.

\begin{deluxetable*}{lcccccccccr}
\tabletypesize{\scriptsize}
\tablecaption{Calculated Parameters}
\tablewidth{0pt}
\tablehead{
\multirow{2}{*}{Parameter~(Unit)}
&\multicolumn{8}{c}{Epoch (Model)}\\
\colhead{\quad}
& \colhead{A ($J=2$)}
& \colhead{A ($J=27$)}
& \colhead{B ($J=2$)}
& \colhead{B ($J=27$)}
& \colhead{C ($J=2$)}
& \colhead{C ($J=27$)}
& \colhead{D ($J=2$)}
& \colhead{D ($J=27$)}
}
\startdata

\\
$r_{\rm Ly\alpha}$ (cm)
&$8.0 \times 10^{16}$
&$8.8 \times 10^{16}$
&$8.3 \times 10^{16}$
&$6.5 \times 10^{16}$
&$4.9 \times 10^{16}$
&$6.5 \times 10^{16}$
&$7.4 \times 10^{16}$
&$7.4 \times 10^{16}$
\\
$r_{\rm H\beta}$ (cm)
&$3.0 \times 10^{17}$
&$3.3 \times 10^{17}$
&$3.1 \times 10^{17}$
&$2.4 \times 10^{17}$
&$1.8 \times 10^{17}$
&$2.4 \times 10^{17}$
&$2.8 \times 10^{17}$
&$2.8 \times 10^{17}$
\\
$R^{\prime}_{\rm blob}$ (cm)
&$5.9  \times 10^{15}$
&$5.5  \times 10^{15}$
&$1.1  \times 10^{16}$
&$5.1  \times 10^{15}$
&$1.8  \times 10^{15}$
&$2.4  \times 10^{15}$
&$7.0  \times 10^{16}$
&$7.0  \times 10^{16}$
\\
$R^{\prime}_{b}/r_{\rm blob}=\phi_{j,{\rm min}}$ ($^{\circ}$)
&$2.0$
&$0.56$
&$3.4$
&$0.62$
&$0.77$
&$0.20$
&$22$
&$6.7$
\\
\hline
\\
$P_{B}$ (erg s$^{-1}$)
&$3.6 \times 10^{44}$
&$3.1 \times 10^{44}$
&$2.1 \times 10^{45}$
&$1.6 \times 10^{45}$
&$3.2 \times 10^{44}$
&$3.9 \times 10^{44}$
&$4.3 \times 10^{45}$
&$4.6 \times 10^{45}$
\\
$P_{e}$ (erg s$^{-1}$)
&$1.5 \times 10^{46}$
&$1.6 \times 10^{46}$
&$1.3 \times 10^{46}$
&$5.8 \times 10^{46}$
&$1.2 \times 10^{46}$
&$6.2 \times 10^{45}$
&$4.5 \times 10^{45}$
&$4.5 \times 10^{45}$
\\
$\zeta_e$ 
&42
&52
&6.2
&36
&38
&16
&1.0
&0.98 
\\
$P_{\rm tot}/P_{\rm acc}$
&0.82
&0.72
&0.67
&4.8
&1.6
&0.53
&0.54
&0.56
\\
\hline
\\
$u_{\rm ext}$ (erg cm$^{-3}$)
&$5.1 \times 10^{-4}$
&$4.7 \times 10^{-4}$
&$3.5 \times 10^{-4}$
&$2.4 \times 10^{-4}$
&$5.9 \times 10^{-5}$
&$9.2 \times 10^{-5}$
&$2.7 \times 10^{-4}$
&$2.8 \times 10^{-4}$
\\
$u_{\rm dust}$ (erg cm$^{-3}$)
&$1.3 \times 10^{-4}$
&$1.5 \times 10^{-4}$
&$1.4 \times 10^{-4}$
&$8.3 \times 10^{-5}$
&$9.5 \times 10^{-6}$
&$3.6 \times 10^{-5}$
&$1.8 \times 10^{-4}$
&$1.7 \times 10^{-4}$
\\
$u_{\rm BLR}$ (erg cm$^{-3}$)
&$3.8 \times 10^{-4}$
&$3.2 \times 10^{-4}$
&$2.1 \times 10^{-4}$
&$1.6 \times 10^{-4}$
&$5.0 \times 10^{-5}$
&$5.6 \times 10^{-5}$
&$9.0 \times 10^{-5}$
&$1.2 \times 10^{-4}$
\\
$u_{\rm Ly\alpha}$ (erg cm$^{-3}$)
&$3.8 \times 10^{-4}$
&$7.0 \times 10^{-8}$
&$2.1 \times 10^{-4}$
&$2.2 \times 10^{-8}$
&$5.0 \times 10^{-5}$
&$1.0 \times 10^{-9}$
&$9.0 \times 10^{-5}$
&$1.0 \times 10^{-8}$
\enddata
\label{table4}
\end{deluxetable*}

\begin{figure}
\vspace{2.2mm} 
\centering
\epsscale{1.1} 
\plotone{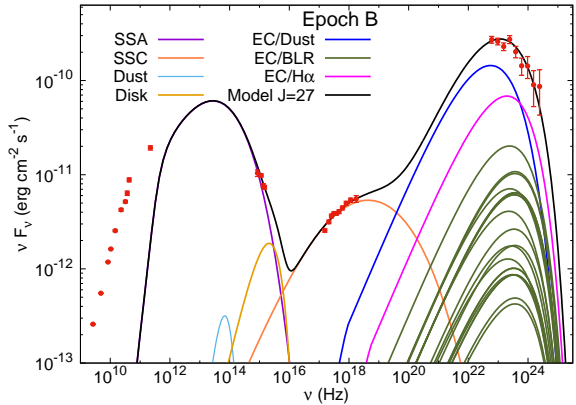}
\caption{Same as Figure~\ref{fig-A_SED}, except the SED data in red are from Epoch B of \citet{hayash12}.}
\label{fig-B_SED}
\vspace{2.2mm}
\end{figure}

Epoch A was a relatively long quiescent period ($\sim$6 weeks) in 2008 Aug-Sept \citep{hayash12}.  In Figure~\ref{fig-A_SED} we qualitatively fit the data from this epoch, using the stratified BLR model with $J=27$. Each of the spectral components is plotted individually for clarity.  The EC of dust photons is the primary contributor to the soft $\g$-rays, but the primary contributor to the hard $\g$-rays is the scattered BLR radiation; more than one EC component is needed to successfully model the $\g$-ray spectrum \citep[see also][]{finke10_3c454}.  The EC for all lines are shown, but we highlight the H$\alpha$ component, which is the most prominent contributor to the $\g$-ray spectrum.  From Table \ref{table3}, one sees that the distance of the blob from the black hole is $r_{\rm blob}=5.55\times10^{17}\,$cm, which is just outside the H$\alpha$ line radius ($r_{\rm H\alpha} \approx 4.3 \times 10^{17}\ \cm > r_{\rm Ly\alpha}$) from Table \ref{table4}.  Despite the fact that Ly$\alpha$ has the highest energy density for $r_{\rm blob} < r_{\rm Ly\alpha}$ (i.e., $u_{\rm line,0}$; recall Equation (\ref{eq-uline})), it does not have the highest energy density at the location of the blob (i.e., $u_{\rm line}$), since the $r_{\rm H\alpha}\approx 5r_{\rm Ly\alpha}$ \citep{finke16}.

Figure \ref{fig-B_SED} depicts each spectral component, and their sum, from our full Compton/stratified BLR model in comparison to multiwavelength data from Epoch B \citep{hayash12,dermer14}, which lasted for 20 days in 2008 Nov-Dec, and comprises a single flare in the optical and $\g$-ray regimes.  Similarly, in Figure \ref{fig-D_SED} we plot each spectral component, and their sum, from our full Compton/stratified BLR model in comparison to multiwavelength data from Epoch D \citep{hayash12,dermer14}, which comprises a single optical/$\gamma$-ray flare that lasted for for 5 days in 2009 Feb.  Neither the Epoch B nor the Epoch D flares were detected in the X-rays, although the X-ray flux was somewhat higher in Epochs B-D than during the quiescent Epoch A.

The spectra plotted in Figures \ref{fig-B_SED} and \ref{fig-D_SED} are qualitatively similar to that plotted in Figure \ref{fig-A_SED}.  However, since they are flares, rather than quiescent spectrum observed in Epoch A, the synchrotron curve dominates over the disk emission in the optical range.  The disk luminosity, as shown in Table \ref{table3}, is fairly constant, perhaps even lower during the Epoch B and D flares as compared with the quiescent Epoch A. In Epoch B (Figure \ref{fig-B_SED}), the X-ray data is independently explained by the SSC component, a consequence of the higher magnetic field, among other contributing parameters, whereas other epochs tend to require EC/dust to produce the harder X-rays.  In both Epochs B and D, as with Epoch A, modeling the $\g$-rays requires both the EC/dust component for soft $\g$-rays, and the EC/BLR for the hard $\g$-rays, and the most important BLR line for EC is H$\alpha$.

For Epoch C (Figure \ref{fig-SEDwBLR}), the result is somewhat different.  Both EC/dust and EC/BLR components are still required to produce the $\g$-ray spectrum, but the most important line is H$\gamma$, at a radius even farther out than H$\alpha$ by a factor of $\sim 2$.  This is consistent with the blob distances $r_{\rm blob}$ in Table \ref{table3}, where the blob in Epoch C is somewhat further from the BH, while the extent of the BLR remains mildly contracted during the entire high flux period, which was subdivided into Epochs B-D, compared to the quiescent Epoch A.

\begin{figure}
\vspace{2.2mm} 
\centering
\epsscale{1.1} 
\plotone{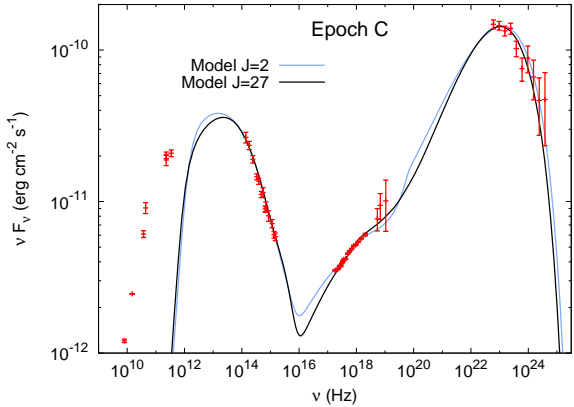}
\caption{SED data in red (from Epoch C of \citet{hayash12}), compared with our steady-state model, computed  using the full BLR implementation ($J=27$; black curve) or only a single line plus dust ($J=2$; blue curve). Note that the model curves are quite similar, despite the fact that the input and calculated parameters differ significantly (see Tables~\ref{table3} and \ref{table4}).}
\label{fig-C_SED}
\vspace{2.2mm}
\end{figure}

In Tables \ref{table3} and \ref{table4} we also report our model parameters and results for the $J=2$ model, i.e., when Ly$\alpha$ is the only line used for Compton scattering, in addition to the dust emission.  A comparison of the $J=2$ and $J=27$ models is carried out in Figure \ref{fig-C_SED}.  The models both reproduce the data well, and indeed the model curves in Figure \ref{fig-C_SED} look very similar.  Note also that the total BLR energy density ($u_{\rm BLR}$) is very similar for both the $J=2$ and $J=27$ cases for all epochs (Table \ref{table4}).  However, as can be seen in Table \ref{table3}, the full (more physically realistic) $J=27$ model has a larger value for the blob distance, $ r_{\rm blob}$, by a factor of $\sim 2-5$.  This indicates that modeling the BLR accurately is important for determining the location of the emitting region.  We emphasize that the $J=27$ model does not have more free parameters than the $J=2$ model, since we have adopted the formalism of \citet{finke16}. Hence, once the disk luminosity has been specified, all of the line luminosities and radii are determined as described in Section \ref{ECblr}.

\subsection{Comparison with Previous Models}

\citet{hayash12} modeled the SEDs for 3C~279 for the same epochs that we analyze here. These authors utilized two models: a propagating emission region model, and a jet precession scenario, which included two synchrotron components.  They chose not to reproduce the X-rays with their models for most of the epochs, since the polarization and variability differences between the X-rays and the $\gamma$-rays suggested that they were produced by different electron populations.  They usually assumed the blob was outside the BLR.  In most of our models, the X-rays come from SSC; the exceptions being the hard X-rays in epochs A and C which have a small contribution from EC.  The SSC can vary with a changing $R\p_b$ independent of the optical (synchrotron) and $\g$-rays (EC) in our model. 

\citet{dermer14} modeled the full SEDs, including the X-rays, but they made a key assumption that differentiates their model from ours: they assumed equipartition among the energy densities in the emitting region.  They chose to reproduce the X-ray emission along with the rest of the IR through $\g$-ray SEDs.  However, they had difficulty explaining the $\g$-ray data at $\ga 5$\ GeV.  Our models which are most similar to the \citet{dermer14} models are our $J=2$ ones, since these only have EC/Ly$\alpha$ and EC/dust, just like \citet{dermer14}.  For these $J=2$ models, we still reproduce the data better than \citet{dermer14}. 

It is important to note that for a given electron distribution, our spectral calculations are identical to those of \citet{dermer14}. Hence the primary difference between our model and theirs is in how the electron distribution is determined. In our model, the electron distribution is determined self-consistently via solution to the electron transport equation (Section \ref{partdist}), rather than using a simple ad hoc electron distribution. Furthermore, we do not assume equipartition.  The assumption of equipartition gives \citet{dermer14} another constraint, and less flexibility, which could account for why our models reproduce the data better. Our model electron distributions more closely resemble power-laws with exponential cutoffs, due to the inclusion of the full Compton cross-section, and the inefficient Compton cooling in the extreme KN regime. These effects allow the electrons to be accelerated to higher energies than they would be able to achieve in a Thomson model \citep[e.g.,][see Figure \ref{fig-EvolutionEps}]{dermer02_KN,moderski05}. Hence the electrons in our model radiate higher energy $\g$-rays, in better agreement with the data.

\citet{asano15} examined Epoch D with a time-dependent stochastic acceleration model in which the electrons are accelerated from some distance $R_0=0.023\,$pc out to $2R_0$, and then radiate.  In this picture, $2R_0 = 1.4 \times 10^{17}$ cm, which is very similar to the blob distance $r_{\rm blob}$ that we find for Epoch D in our $J=2$ model.  This is consistent with their use of the same EC model as \citet{hayash12}, including external radiation from the BLR (as a single source) and dust. Their magnetic field $B$ varies with the blob distance, but when acceleration turns off and radiation turns on, $B=3.5\,$G, which is more than 6 times what we found in Epoch D.  Their Doppler factor $\dD=15$ is somewhat smaller, but comparable to our value of $\dD=20$. Their energy densities are appreciably larger than ours (see Table \ref{table4}), with $u_{B} = 0.48$ erg cm$^{-3}$, and $u_{\rm ext}$ not provided directly, but stated to be much larger.  In contrast, our $u_B = 1.2 \times 10^{-2}$ (erg cm$^{-3}$) for $J=2$ in Epoch D, and $u_{\rm ext}$ is almost 100 times smaller.  So, their jet power must be very strongly field dominated  even though their particle injection rate $\dot{N}_e = 7.8 \times 10^{49}$ s$^{-1}$ of $\gamma_e=10$ electrons is $2 \times 10^{14}$ s$^{-1}$ times higher than our injection of $\g_{\rm inj}=1.01$ electrons.

\subsection{Non-Equipartition Jet}

\citet{dermer14} assumed equipartition between the energy densities of the particles and the magnetic field, in order to constrain their SED modeling of 3C 279.  Equipartition results in jet powers close to the minimum jet power.  We do not assume equipartition in our modeling.  We compute the equipartition parameter $\zeta_e=u_e/u_B=P_e/P_B$, following \citet{dermer14}, and the results are reported in Table \ref{table4}.  Our models are either electron dominated or near equipartition.  Since some of these results are out of equipartition, and hence far from the minimum power condition, we also compute the total jet power ($P_{\rm tot} = P_e + P_B$) as a fraction of accretion power ($P_{\rm acc}=L_{\rm disk}/0.4$, where the 0.4 is the efficiency for a maximally-rotating black hole), and these results are also reported in Table \ref{table4}.  In all but two of our models, we find that $P_{\rm tot}/P_{\rm acc}<1$, indicating that accretion is sufficient to explain the power in the jet.  General relativistic magnetohydrodynamic simulations carried out by \citet{tchek11} indicate that $L_{\rm tot}/P_{\rm acc} \sim $\ a few is possible with magnetically arrested accretion, which helps to support our high values.  In the case of a spinning BH, significant energy can be drawn out of the ergosphere via the Blandford-Znajek process, reducing the need to power the blazar jet using accretion alone.

\subsection{Electron Number and Energy Budgets}

\begin{deluxetable*}{lcccccccr}
\tabletypesize{\scriptsize}
\tablecaption{Electron Energy Budget in Co-moving Blob Frame}
\tablewidth{\textwidth}
\tablehead{
Parameter
&\multicolumn{8}{c}{Epoch (Model)}\\
\colhead{(Unit)}  
& \colhead{A ($J=2$)}
& \colhead{A ($J=27$)}
& \colhead{B ($J=2$)}
& \colhead{B ($J=27$)}
& \colhead{C ($J=2$)}
& \colhead{C ($J=27$)}
& \colhead{D ($J=2$)}
& \colhead{D ($J=27$)}
}
\startdata
(s$^{-1}$)
\\  
\vspace{0.1cm}
$\dot{N}'_{\rm inj}$ 
&$\quad 8.90\times10^{35}$
&$\quad 8.80 \times 10^{35}$
&$\quad 2.10 \times 10^{35}$
&$\quad 1.35 \times 10^{35}$
&$\quad 8.22 \times 10^{35}$
&$\quad 5.00 \times 10^{35}$
&$\quad 4.00 \times 10^{35}$
&$\quad 3.85 \times 10^{35}$
\\
$\dot{N}'_{\rm esc}$ 
&$-8.90 \times 10^{35}$
&$-8.80 \times 10^{35}$
&$-2.10 \times 10^{35}$
&$-1.35 \times 10^{35}$
&$-8.22 \times 10^{35}$
&$-5.00 \times 10^{35}$
&$-4.00 \times 10^{35}$
&$-3.85 \times 10^{35}$
\vspace{0.1cm}
\\
\hline
\vspace{0.1cm}
(erg s$^{-1}$)
\\
$P^{\prime}_{\rm inj}$ 
&$\quad 7.36 \times 10^{29}$
&$\quad 7.28 \times 10^{29}$
&$\quad 1.74 \times 10^{29}$
&$\quad 1.12 \times 10^{29}$
&$\quad 6.80 \times 10^{29}$
&$\quad 4.14 \times 10^{29}$
&$\quad 3.31 \times 10^{29}$
&$\quad 3.18 \times 10^{29}$
\\
$P^{\prime}_{\rm esc}$ 
&$-4.21 \times 10^{31}$
&$-4.51 \times 10^{31}$
&$-2.35 \times 10^{31}$
&$-1.35 \times 10^{31}$
&$-2.31 \times 10^{30}$
&$-1.24 \times 10^{31}$
&$-7.36 \times 10^{31}$
&$-6.81 \times 10^{31}$
\\
$P^{\prime}_{\rm sto}$ 
&$\quad 7.64 \times 10^{42}$
&$\quad 7.95 \times 10^{42}$
&$\quad 8.00 \times 10^{42}$
&$\quad 1.67 \times 10^{42}$
&$\quad 4.23 \times 10^{41}$
&$\quad 8.28 \times 10^{41}$
&$\quad 1.50 \times 10^{44}$
&$\quad 1.54 \times 10^{44}$
\\
$P^{\prime}_{\rm sh,ad}$ 
&$-7.26 \times 10^{42}$
&$-7.55 \times 10^{42}$
&$-7.20 \times 10^{42}$
&$-1.46 \times 10^{42}$
&$-4.28 \times 10^{41}$
&$-8.08 \times 10^{41}$
&$-1.39 \times 10^{44}$
&$-1.42 \times 10^{44}$
\\
$P^{\prime}_{\rm syn}$ 
&$-6.97 \times 10^{40}$
&$-6.60 \times 10^{40}$
&$-1.31 \times 10^{41}$
&$-3.98 \times 10^{40}$
&$-3.96 \times 10^{38}$
&$-4.62 \times 10^{39}$
&$-1.45 \times 10^{42}$
&$-1.49 \times 10^{42}$
\\
$P^{\prime}_{\rm EC}$ 
&$-3.14 \times 10^{41}$
&$-3.33 \times 10^{41}$
&$-6.70 \times 10^{41}$
&$-1.69 \times 10^{41}$
&$-1.31 \times 10^{39}$
&$-1.74 \times 10^{40}$
&$-9.85 \times 10^{42}$
&$-1.00 \times 10^{43}$
\\
\\ 
\vspace{0.1cm}
$P^{\prime}_{\rm net}$ 
&$-1.43 \times 10^{39}$
&$-1.62 \times 10^{39}$
&$-5.82 \times 10^{38}$
&$-1.19 \times 10^{38}$
&$-6.99 \times 10^{39}$
&$-1.31 \times 10^{39}$
&$-1.57 \times 10^{40}$
&$-1.55 \times 10^{40}$
\\ 
\hline 
\vspace{-0.2cm}
\\
$\delta_{\rm err}$
&$0.02\%$
&$0.02\%$
&$0.01\%$
&$0.01\%$
&$1.65\%$
&$0.16\%$
&$0.01\%$
&$0.01\%$
\vspace{0.01cm}
\enddata
\label{table5}
\end{deluxetable*}

One of the strengths of utilizing a particle transport equation such as the one employed here (Equation~(\ref{eq-transport})) is that the formalism automatically conserves energy. This provides us with a unique means for gaining further insight into how the energy is distributed among the various available channels associated with the acceleration, loss, radiative, and escape processes experienced by the electrons in the blob. We can use this information to explore the power distribution for each of the epochs of the 3C~279 data we analyze here, and to determine what patterns may exist in the data that could provide significant clues about the relevant underlying physical processes. Furthermore, calculating the power in each channel also allows us to confirm that energy is conserved in the model, which lends further support to our conclusions. It is also interesting to confirm that the rate of electron escape from the blob is equal to the injection rate, which is non-trivial in our application since the escape timescale is energy-dependent (see Equation~(\ref{eq-esc})).

The steady-state transport equation with the full Compton cross-section implemented was solved using a numerical procedure, as described in Section \ref{solnmeth} and Appendix \ref{ap-norm}. The rate of escape of electrons from the blob, measured in the blob frame, is computed using
\begin{equation}
\dot{N}\p_{\rm esc} = \int_{\gp_{\rm min}}^{\gp_{\rm max}} \frac{D_0 \gp}{\tau} N\p_e(\gp) d\gp \ ,
\label{eq-esc-rate}
\end{equation}
where the primes in this section emphasize that the powers, injection rates, and escape rates considered here are calculated in the co-moving (blob) frame, whereas the jet powers shown in Table \ref{table4} are given in the stationary (BH) frame. The integral in Equation~(\ref{eq-esc-rate}) is computed using the full numerical solution for the electron distribution $N\p_e(\g')$. In Table \ref{table5}, we compare the results obtained for $\dot N'_{\rm esc}$ with the electron injection rate, $\dot{N}'_{\rm inj}$, and we note that for each epoch analyzed, we obtain identical values for these two quantities. This confirms that the electron ``number budget'' is correctly handled in our formalism. The particle injection rate is related to the injection luminosity $L'_{\rm e,inj}$ by Equation (\ref{eq-injLum}).

In addition to tracking the particle number budget, we can also calculate the power produced by each individual process in the transport equation in order to verify the energy budget in our numerical routine balances properly, and to compare the relative energy contributions of each process affecting the electron distribution in the blob.  The power provided by incident particles is the same as the injection luminosity ($P\p_{\rm inj} = L'_{\rm e,inj}$).  The escape power is given by
\begin{equation}
P\p_{\rm esc} = m_e c^2 \frac{D_0}{\tau} \int_{\gp_{\rm min}}^{\gp_{\rm max}} {\gp}^2 N\p_e(\gp) d\gp \ .
\end{equation}

The remaining components of the electron power budget are calculated by integrating the associated terms in the Fokker-Planck drift coefficient (Equation~(\ref{eq-driftKN})), since the broadening coefficient does not contribute to the energy budget.  The powers associated with stochastic acceleration, shock acceleration+adiabatic losses, synchrotron losses, and inverse-Compton (EC) losses are denoted by $P\p_{\rm sto}$, $P\p_{\rm sh,ad}$, $P\p_{\rm syn}$, and $P\p_{\rm EC}$, respectively. Each of these quantities can be computed by integrating the respective term in the drift coefficient using the template
\begin{equation}
P\p_{\rm drift} = m_ec^2 D_0 \int_{\gp_{\rm min}}^{\gp_{\rm max}} C_{\rm drift}(\gp) N\p_e(\gp) d\gp \ ,
\end{equation}
where 
\begin{equation}
C_{\rm drift}(\gp) = \begin{cases}
4\gp \\
a \gp \\
-b_{\rm syn} {\gp}^2 \\
-b_{\rm C} {\gp}^2 
\end{cases} \ , 
\end{equation}
in turn for the cases of stochastic, shock/adiabatic, synchrotron, and Compton losses, respectively.  The results obtained for each of these power components (in the blob frame) for each model and epoch of 3C 279 treated here are reported in Table \ref{table5}. We also compute the sum of the gain and loss terms, given by
\begin{equation}
P\p_{\rm net} = P\p_{\rm inj} + P\p_{\rm esc} + P\p_{\rm sto} + P\p_{\rm sh,ad} + P\p_{\rm syn} + P\p_{\rm EC} \ .
\end{equation}
Ideally, we should obtain $P\p_{\rm net}=0$ if energy is perfectly conserved in our model. However, due to numerical errors, this can never be the case. We therefore compute a relative error, $\delta_{\rm err}$, by dividing $| P\p_{\rm net} |$ by the sum of the stochastic acceleration rate, $P\p_{\rm sto}$ and the injection rate $P\p_{\rm inj}$, which we report in Table~\ref{table5}. We confirm that the relative error is quite small in most cases, which is sufficient to confirm that energy is properly conserved in our model applications to each epoch of the 3C~279 data treated here.

Comparing the various channel powers listed in Table \ref{table5}, we are able to draw some interesting conclusions about the physical processes influencing the behavior of the electrons in the blob.  The particle injection power is always more than 10 orders of magnitude lower than the rest of the positive power contributions because the rate of particle injection is very low.  As a result, the steady-state electron energy distributions we obtain are insensitive to the precise energy (or distribution) of the injected particles. The lack of significant pre-acceleration of the electrons is consistent with the observed lack of high-energy radiation along the upstream path of the jet, which implies that the precursor jet is ``cold.'' In this scenario, the energy of the electrons in the precursor jet is mostly in the form of directed (bulk) kinetic energy, rather than stochastic energy. This situation reverses once the electrons are energized in the acceleration region. We therefore conclude that the acceleration zone is located coincident with, or at least nearby, the emission region, which also helps to motivate the one-zone model considered here.

The stochastic acceleration power and the combined shock acceleration and adiabatic expansion power are the largest contributors to the overall energy budget.  Stochastic acceleration provides the bulk of the acceleration to the electrons in the blob, while adiabatic expansion is the largest energy loss mechanism.  This is true for all epochs of 3C 279 that we examined, as can be seen in Table \ref{table5}.  The shock acceleration  and adiabatic expansion processes are combined into single term in our model, represented by the constant $a$ in the transport equation, because they have the same first-order dependence on the particle energy. In general, adiabatic losses are expected to lead to the value $a \sim -4$ \citep{lewis16}. Since the values for $a$ reported in Table \ref{table3} are all close to $-4$, this suggests that shock acceleration (which always makes a positive contribution to $a$) is a weak contributor to the acceleration of the electrons. Hence we conclude that stochastic acceleration is the main contributor to the acceleration of the electrons in the blob.

Comparison of our model predictions with the observational data leads to the conclusion that stochastic acceleration and adiabatic expansion are the dominant processes in the examined epochs of 3C 279 (due to the breadth of the individual spectral components), as discussed in Section \ref{paramstudy}, and  depicted in Figure \ref{fig-ParamStdyMod}. The active region, where the particle acceleration and emission takes place, is located $\sim 0.1\,$pc from the central engine. We note that the generation of high-energy emission at this distance is consistent with Fermi observations \citep[e.g.][]{madejski16}. The occurrence of strong adiabatic cooling $\sim 0.1\,$pc from the black hole would imply substantial local acceleration of the bulk flow, which is supported by the observations \citep[e.g.][]{homan15}. The acceleration could be due to passage through a recollimation shock that creates a nozzle in the flow \citep[e.g.][]{cohen15}. The simultaneous dominance of stochastic particle acceleration at the same location may reflect the action of wave excitation at the recollimation shock \citep{marscher14}. Taken together, these concepts support the scenario explored in the present paper, in which adiabatic losses and stochastic acceleration both dominate at a distance of $\sim 0.1\,$pc during the examined epochs of 3C 279.

The energy losses to the electrons due to the EC process are always greater in magnitude (more negative) than those due to the synchrotron process for all epochs of 3C 279 considered here.  This is expected, because the part of the radiation spectrum due to EC always displays a higher amplitude in the $\nu F_\nu$ spectrum than that of the synchrotron component for all of the 3C 279 epochs we examined.
The electron energy losses due to radiation processes (EC and synchrotron) are always less than the losses due to adiabatic expansion.

\section{Discussion}
\label{discuss}

We have created a new one-zone leptonic model for particle acceleration and emission in blazar jets.  We treat electron acceleration, escape, and losses realistically and self-consistently with a steady-state Fokker-Planck equation, which includes the full Compton energy loss term, valid in both the Thomson and Klein-Nishina regimes.  We then use the resulting electron distribution to compute the emission expected for the FSRQ 3C~279, including Compton scattering of all relevant broad emission lines.  We show that the model can reproduce Epochs A-D of the 2008-2009 spectral data for 3C 279 reported by \citet{hayash12}, including the hard $\gamma$-ray component. We note that our method is somewhat more general than that applied in most previous studies of blazar emission properties, since we make fewer simplifying assumptions about the acceleration and losses processes. We compare our model and our astrophysical results with those obtained by other authors below.

\subsection{Comparison with Previous Work}

\citet{diltz14} studied electron evolution and emission in blazars using a Fokker-Planck equation that included second-order Fermi acceleration processes.  However, they did not include a term representing first-order Fermi acceleration. 
Furthermore, they treated the BLR as a blackbody at a particular radius, not taking into account the various broad lines at different radii.

\citet{asano14} developed a particle acceleration and emission model that is somewhat similar to ours, although they considered Kolmogorov and Bohm stochastic acceleration, in addition to the hard-sphere prescription that we implement here. \citet{asano15} applied the time-dependent model of \citet{asano14} to study FSRQs flares, and they also used the steady-state model to reproduce epoch D of 3C 279 from \citet{hayash12}. In the time-dependent model of \citet{asano15}, the electrons are injected into a blob, and then stochastically accelerated until the blob location reaches a predetermined distance from the BH, at which point the acceleration is turned off and the radiative emission processes take over. They include synchrotron, SSC, EC/dust, and EC/BLR as their radiative processes, and they also include adiabatic energy losses.

It is interesting to highlight the differences between the treatment of the BLR seed photons for the EC process considered here, versus previous treatments. We employ a radially stratified model in which 26 individual species of seed photons are emitted from separate, discrete radial shells. \citet{hayash12} and \citet{asano15} employ a similar discrete shell geometry, except that they consider only one ``equivalent'' line component, instead of utilizing an explicit multi-component BLR emission model like ours. In their approach, the equivalent line is a conceptual average of the multiple seed photon lines known to exist in the environment surrounding the BH. \citet{asano14} employ an isotropic (non-stratified) shell geometry. In order to justify the geometrical assumption in their model, the blob must be located inside the BLR, but there is no such restriction in our model.

The electron acceleration in the model of \citet{asano15} is purely stochastic, and there is no explicit treatment of shock acceleration. Instead, shock acceleration is implemented approximately by invoking a population of pre-accelerated electrons with a prescribed energy $\g_{\rm inj}=100$ that is much higher than ours \citep[see also][]{asano11}. We improve upon the model of \citet{asano15} by including shock acceleration explicitly, as well as an energy-dependent Bohm particle escape term.

\citet{diltz14} numerically solved a time-dependent transport equation, including synchrotron and Compton energy losses, hard-sphere stochastic acceleration, and energy-independent particle escape. They focused on the injection of electrons with a power-law energy distribution, which was intended to model the pre-acceleration of the electrons due to encounters with shocks in the jet. They used both the BLR and dust torus emission as seed photon sources for EC, but they modeled the BLR as a single blackbody at a single radius, whereas we employ a detailed stratified model for the BLR. They also assumed that the EC seed photon distribution is isotropic, and consequently their physical interpretation is constrained to the case where the blob is located inside the BLR radius. Conversely, our model accommodates blob locations both in and outside of the BLR. The \citet{diltz14} model fits their sample dataset well, and they are able to conclude that a decrease in the overall acceleration timescale is related to a correlation between the optical and $\g$-ray bands, which was observed during several 3C 279 flares.

\citet{chen14} used a multizone inhomogeneous and time-dependent model to investigate FSRQ emission regions.  They implement continuous stochastic acceleration, radiative cooling, particle escape, and the injection of a pre-accelerated electron distribution. Thus, shock acceleration was not included in their transport equation. Their treatment of the EC process assumes that the incident radiation is generated in the dust torus only, and hence they neglect the BLR contribution. Their results suggest that different types of flares are associated with different magnetic field orientations. They conclude that SSC models are not preferred for FSRQ, in agreement with our findings.

\citet{tramacere11} also studied blazar flares using a stochastic acceleration transport equation, including synchrotron and EC energy losses. They assumed that the electrons were pre-accelerated before experiencing stochastic acceleration. They argue for a log-parabola electron distribution based on the dominance of stochastic acceleration \citep[see also][]{dermer14}. In our work, the injected electrons have a very low Lorentz factor, $\gamma_{\rm inj} = 1.01$, in order to simulate the effect of thermal electron injection, without having to resort to electron pre-acceleration. In this sense, our model is more self-consistent than many previous works.

\subsection{Summary}

We have developed a novel calculation of the electron distribution describing the electrons in a blob propagating through a blazar jet. The model is based on a steady-state electron transport equation. After solving the transport equation, we used the resulting electron distributions to self-consistently calculate the multiwavelength spectrum, using either 2 and 27 sources of seed photons for the EC process. We applied our model to interpret and approximately fit the 2008-2009 spectral data for 3C 279, reported by \citet{hayash12} and further analyzed by \citet{dermer14}.

Our inclusion of two acceleration mechanisms in the electron transport equation distinguishes our work from previous models because the shape of the electron distribution is allowed to vary as different processes become dominant. Hence the solutions we obtain for the electron distribution improve upon the previous ad hoc distributions, and they also improve upon those obtained by solving a transport equation that only includes one form of acceleration. For example, a transport equation where only stochastic acceleration is present will always produce a log-parabola electron distribution \citep{tramacere11}. Models that do not include shock acceleration explicitly usually assume that the injected electrons are pre-accelerated \citep[e.g.][]{tramacere11,diltz14,chen14,asano15}. In contrast, we do not assume particle pre-acceleration, and instead treat all acceleration processes explicitly using the transport equation. Hence our model is a more fundamental, first-principles representation of the physical processes occurring in the blob. Our primary findings are listed below:

\begin{itemize}
\item A one-zone leptonic model is sufficient to match the data in all epochs examined.

\item We include shock and stochastic acceleration in our model, and thereby avoid invoking the injection of pre-accelerated electrons. Instead, all of the particle acceleration processes are treated self-consistently using a rigorous transport equation.

\item Stochastic acceleration dominates over shock acceleration in all of the calculations performed here.

\item The distance of the blob from the BH is dependent on the structure and composition of the BLR.  Specifically, the blob is calculated to be further from the BH when the BLR is stratified.

\item In our analysis of Epochs A-D from \citet{hayash12}, shock acceleration is present, but overwhelmed by adiabatic losses.

\item The utilization of a detailed, stratified model for the BLR leads to improved estimates for the model parameters.

\item We calculate the the matter and magnetic energy densities, and conclude that the jet is not always in equipartition, but sometimes matter dominated.
\end{itemize}

We plan to use the new model developed here to analyze data for additional blazar $\gamma$-ray flares in future work.

\acknowledgements

We thank the anonymous referee for insightful comments, which improved the presentation of the manuscript.
T.R.L was supported as a summer intern at NRL through a NASA contract S-15633Y.  J.D.F. was supported by the Chief of Naval Research.

\appendix

\section{Numerical Solution of the Electron Transport Equation} 
\label{ap-norm}

Here we review two novel methodologies developed in the course of this work that may be useful in the context of other situations involving the numerical solution to stiff ordinary differential equations.

\subsection{Boundary Conditions}

When the full Compton cross-section is utilized in the transport equation (Equation~(\ref{eq-transport})), an analytical solution is no longer possible. We therefore solve this equation using a bi-directional Runge-Kutta method. This approach requires two separate integrations of the transport equation, with one starting at the high-energy boundary, $\gamma = \gamma_{\rm max}$, and the other starting at the low-energy boundary, $\gamma = \gamma_{\rm min}$. In order to apply this method, we therefore need to specify appropriate boundary conditions at the two ends of the computational domain. We begin by writing Equation~(\ref{eq-transport}) in the flux-conservation form
\begin{equation}
\frac{1}{D_0} \frac{\partial N_e}{\partial t} = 0 = -\frac{1}{D_0} \frac{\partial \dot{\mathcal{N}}}{\partial \gamma} - \frac{N_e \gamma}{\tau} + \frac{\dot{N}_{\rm inj}\delta(\gamma-\gamma_{\rm inj})}{D_0} \ ,
\label{eq-teq-fcon}
\end{equation}
where the electron transport rate in the energy space is defined by
\begin{equation}
\dot{\mathcal{N}} \equiv D_0 \bigg\{ -\gamma^2 \frac{\partial N_e}{\partial \gamma} + \big[2\gamma
+ a \gamma - b_{\rm syn} \gamma^2
- \gamma^2 \sum_{j=1}^J b_{\rm C}^{(j)} H(\g\e^{(j)}_{\rm ph}) \big] N_e \bigg\} \ .
\label{eq-flux}
\end{equation}
At very low energies ($\gamma \to 0$), the electron transport rate, $\dot{\mathcal{N}}$, must vanish, since no particles can be transported into the negative energy domain. Likewise, at sufficiently high energies, we expect that $\dot{\mathcal{N}} \to 0$ because the electrons cannot be accelerated to arbitrarily high energies.

The maximum energy achieved by the electrons will be comparable to the exponential turnover energy in the electron distribution, $N_e(\gamma)$, which corresponds to a balance between acceleration and losses. This energy, which we refer to as the ``equilibrium energy,'' $\gamma_{\rm eq}$, can be estimated by setting the Fokker-Planck drift coefficient (Equation~(\ref{eq-driftKN})) equal to zero. At sufficiently high electron energies, the Compton cross-section declines exponentially due to the KN effect, and therefore synchrotron losses will dominate. The equilibrium energy is therefore given by \citep{lewis16}
\begin{equation}
\gamma_{\rm eq} = \frac{4 + a}{b_{\rm syn}} \ .
\label{eq-gamma-eq}
\end{equation}
In general, we find that $\gamma_{\rm eq}$ exceeds the maximum Lorentz factor in our computational domain, $\gamma_{\rm max}$, and therefore we are justified in applying the zero-flux boundary condition at both the high- and low-energy boundaries in our Runge-Kutta code. Our zero-flux boundary condition can therefore be written as
\begin{equation}
\dot{\mathcal{N}}(\gamma) = 0 \ , \qquad
\begin{cases}
\gamma = \gamma_{\rm min} \ , \\
\gamma = \gamma_{\rm max} \ .
\end{cases}
\label{eq-flux2}
\end{equation}
Combining Equations~(\ref{eq-flux}) and (\ref{eq-flux2}) yields an expression for the derivative $d N_e/d\gamma$, which can be written as
\begin{equation}
\frac{\partial N_e}{\partial \g} = \bigg[ \frac{2 + a}{\g} - b_{\rm syn} -\sum_{j=1}^J b_{\rm C}^{(j)} H(\g\e^{(j)}_{\rm ph}) \bigg] N_e(\g) \ .
\label{eq-flux-deriv}
\end{equation}
Equation~(\ref{eq-flux-deriv}) is applied at $\gamma = \gamma_{\rm min}$ and also at $\gamma = \gamma_{\rm max}$ to calculate the starting derivative, once a boundary value for $N_e$ has been specified, as discussed below.

\subsection{Normalization Method}

One of the challenges of the Runge-Kutta implementation is that the equation is mathematically stiff, and it therefore requires extremely high grid resolution in order to obtain sufficient accuracy using standard methods. We therefore develop a novel alternative approach here that is based on ensuring a proper normalization for the global electron energy distribution, $N_e(\gamma)$.

The general solution for the electron energy distribution in our application is obtained by numerically integrating the homogeneous version of Equation~(\ref{eq-teq-fcon}) obtained when $\gamma \ne \gamma_{\rm inj}$, so that the source term is not active. We use a bi-directional technique to integrate Equation~(\ref{eq-teq-fcon}) away from the high-energy boundary at $\gamma = \gamma_{\rm max}$, and likewise away from the low-energy boundary at $\gamma = \gamma_{\rm min}$. The two fundamental solutions for $N_e$ in the regions $\gamma \le \gamma_{\rm inj}$ and $\gamma \ge \gamma_{\rm inj}$ are denoted by $f_{\rm inc}$ and $f_{\rm dec}$, which refer to as the ``incrementing'' and ``decrementing'' solutions, respectively. The derivatives at the two respective boundaries are each set using Equation~(\ref{eq-flux-deriv}). We set $f_{\rm inj}=1$ at $\gamma = \gamma_{\rm min}$ and $f_{\rm dec} = 1$ at $\gamma = \gamma_{\rm max}$ without loss of generality, since the global solution for $N_e(\gamma)$ will be renormalized later in the process.

Once the fundamental solutions $f_{\rm inc}$ and $f_{\rm dec}$ have been obtained, the normalized global solution for the electron number distribution is given by 
\begin{equation}
N_e(\gamma) = \begin{cases} 
A f_{\rm inc}(\gamma) \ , \quad \gamma \le \gamma_{\rm inj} \ , \\
B f_{\rm dec}(\gamma) \ , \quad \gamma \ge \gamma_{\rm inj} \ ,
\end{cases}
\end{equation}
where $A$ and $B$ are normalization coefficients, that can be computed by ensuring a steady-state balance between electron injection and escape. The rate at which electrons escape from the blob is given by
\begin{equation}
\dot{N}_{\rm esc} = \int_0^{\infty} \frac{N_e(\gamma)}{t_{\rm esc}(\gamma)} d\gamma \ ,
\label{eq-esc-appen}
\end{equation}
and the escape timescale, $t_{\rm esc}$, is related to the dimensionless parameter $\tau$ via \citep{lewis16}
\begin{equation}
\frac{1}{t_{\rm esc}(\gamma)} = \frac{\gamma D_0}{\tau} \ .
\end{equation}

In a steady state, the rate of escape of electrons from the blob, $\dot{N}_{\rm esc}$, must equal the injection rate, $\dot{N}_{\rm inj}$, computed using Equation~(\ref{eq-esc-appen}). Hence, we require that
\begin{equation}
\dot{N}_{\rm esc} = \dot{N}_{\rm inj} \ .
\end{equation}
Combining relations, we find that
\begin{equation}
\dot{N}_{0} = A \int_0^{\gamma_{\rm inj}} \frac{f_{\rm inc}(\gamma)}{t_{\rm esc}(\gamma)} d\gamma + B \int_{\gamma_{\rm inj}}^{\infty} \frac{f_{\rm dec}(\gamma)}{t_{\rm esc}(\gamma)} d\gamma \ .
\label{eq-A1}
\end{equation}
The electron distribution must be continuous at the injection energy, therefore
\begin{equation}
A f_{\rm inc}(\gamma_{\rm inj}) = B f_{\rm dec}(\gamma_{\rm inj}) \ .
\end{equation}
Solving for $B$ yields, 
\begin{equation}
B = A \, \frac{f_{\rm inc}(\gamma_{\rm inj})}{f_{\rm dec}(\gamma_{\rm inj})} \ .
\label{eq-A2}
\end{equation}
We can eliminate $B$ between Equations~(\ref{eq-A1}) and (\ref{eq-A2}) and solve for $A$ to obtain
\begin{equation}
A= \dot{N}_{0} \bigg[ \int_0^{\gamma_{\rm inj}} \frac{ f_{\rm inc}(\gamma)}{t_{\rm esc}(\gamma)} d\gamma + \frac{ f_{\rm inc}(\gamma_{\rm inj})}{f_{\rm dec}(\gamma_{\rm inj})} \int_{\gamma_{\rm inj}}^{\infty} \frac{f_{\rm dec}(\gamma)}{t_{\rm esc}(\gamma)} d\gamma \bigg]^{-1} \ .
\end{equation}
The fully normalized global solution for the electron distribution, valid over the entire energy range, $\gamma_{\rm min} \le \gamma \le \gamma_{\rm max}$, can now be written as
\begin{equation}
N_e(\gamma) = A f_{\rm inc}(\gamma_{\rm inj}) \begin{cases}
f_{\rm inc}(\gamma)/f_{\rm inc}(\gamma_{\rm inj}) \ , &\gamma \le \gamma_{\rm inj} \ , \\
f_{\rm dec}(\gamma)/f_{\rm dec}(\gamma_{\rm inj}) \ , &\gamma \ge \gamma_{\rm inj} \ ,
\end{cases}
\label{eq-6}
\end{equation}
We employ a trapezoid numerical integration scheme to compute the constant $A$ in Equation (25) based on the fundamental solutions $f_{\rm inc}$ and $f_{\rm dec}$.

The normalization technique developed here is theoretically equivalent to the usual Wronskian method, but the latter method requires very accurate numerical computation of the Wronskian of the fundamental solutions $f_{\rm inc}$ and $f_{\rm dec}$. Our alternative normalization technique is much more accurate and efficient when dealing with stiff ODEs such as our transport equation.

\clearpage


\bibliographystyle{apj}
\bibliography{variability_ref,EBL_ref,references,mypapers_ref,blazar_ref,sequence_ref,SSC_ref,LAT_ref,3c454.3_ref,msNote}

\end{document}